\documentclass{article}

\usepackage[margin=1in]{geometry}

\usepackage{amsfonts}
\usepackage{setspace}
\usepackage{subfig}
\usepackage{bbm}
\usepackage{epsfig}  
\usepackage{graphicx}  
\usepackage{amsmath}  
\usepackage{amssymb}  
\usepackage{url}  
\usepackage{lscape}  
\usepackage{bbm}
\usepackage{stfloats}
\usepackage{float}
\usepackage{units}
\usepackage[boxed,scleft]{algorithm2e}
\usepackage{enumerate}
\usepackage{comment}
\usepackage{flushend}
\usepackage{import}
\usepackage{color}

\usepackage{array}
\usepackage{url}

\newcommand\Mark[1]{\textsuperscript#1}

\newtheorem{theorem}{Theorem}[section]

\newtheorem{lemma}[theorem]{Lemma}
\newtheorem{definition}[theorem]{Definition}
\newtheorem{corollary}[theorem]{Corollary}
\newtheorem{remark}[theorem]{Remark}

\graphicspath{{./figures/}}
\DeclareGraphicsExtensions{.pdf,.png,.jpg}
\DeclareMathOperator*{\diag}{diag}

\newcommand*{\unity}{\textrm{{\usefont{U}{fplmbb}{m}{n}1}}}

\newcommand{\s}{S}

\newcommand{\bb}{\theta}

\newcommand{\cY}{{\cal Y}}
\newcommand{\cP}{{\cal P}}
\newcommand{\cM}{{\cal M}}

\newcommand{\cT}{{\cal T}}
\newcommand{\prob}{{\mathbb P}}
\newcommand{\hf}{{\hat{f}}}
\newcommand{\ind}{{\mathbb  I}}
\newcommand{\E}{{\mathbb  E}}
\newcommand{\reals}{{\mathbb  R}}
\newcommand{\la}{\langle}
\newcommand{\ra}{\rangle}

\newcommand{\aveacc}{{\rm ACC_{ave}}}
\newcommand{\wcacc}{{\rm ACC_{wc}}}
\newcommand{\w}{w}
\newcommand{\tX}{\tilde{X}}
\newcommand{\tx}{\tilde{x}}
\newcommand{\cR}{{\cal R}}
\newcommand{\Z}{{\mathbb Z}}

\begin{document}

%
%
%
%

\begingroup
\centering
	{\LARGE Differentially Private Multi-party Computation:\\ 
	Optimality of Non-Interactive Randomized Response \\[1.5em]
\large Peter Kairouz \Mark{1},Sewoong Oh \Mark{2}, Pramod Viswanath\Mark{3}}\\[1em]
\begin{tabular}{*{3}{>{\centering}p{.25\textwidth}}}
\Mark{1}Department of Electrical and Computer Engineering  & \Mark{2} Department of Industrial and Enterprise Systems Engineering& \Mark{3}Department of Electrical and Computer Engineering \tabularnewline
University of Illinois at Urbana-Champaign  & University of Illinois at Urbana-Champaign & University of Illinois at Urbana-Champaign \tabularnewline
\url{kairouz2@illinois.edu} & \url{swoh@illinois.edu} & \url{pramodv@illinois.edu}
\end{tabular}\par
\endgroup

\begin{abstract}
We study the problem of interactive function computation by multiple
parties possessing a single bit each in a differential privacy setting
(i.e., there remains an uncertainty in any specific party's bit even when
given the transcript of the interactions and all the other parties' bits).
Each party is interested in computing a function, which could differ from
party to party, and there could be a central observer interested in
computing a separate function. Performance at each party and the central
observer is measured via the accuracy of the function computed. We allow
for an arbitrary cost function to measure the distortion between the true
and the computed function value. Our main result is the exact optimality of
a simple non-interactive protocol: each party randomizes (sufficiently) and
publishes its own bit. In other words, non-interactive randomized response
is exactly optimal. Each party and the central observer then separately
compute their respective function to maximize the appropriate notion of
their accuracy measure. The optimality is very general: it holds for all
types of functions, heterogeneous privacy conditions on  the parties, all
types of cost metrics, and both average and worst-case (over the inputs)
measures of accuracy. Finally,  the optimality result is simultaneous, in
terms of maximizing accuracy at each of the parties and the central
observer.
\end{abstract}

\newpage

\section{Introduction}
\label{sec:intro}



Multi-party computation (MPC) is a generic framework where multiple parties share 
their information interactively amongst themselves towards a goal of computing some function (potentially 
different at each of the parties) of the information.  The challenges are in computing the functions efficiently (to minimize the {\em communication complexity}) and/or in computing 
the functions such that parties learn nothing more about the
 others' information than can be learnt from the output of the function computed (this topic is studied under the rubric of {\em secure function evaluation} (SFE)). 
 These are classical topics:  state of the art of the 
communication complexity formulation is archived in \cite{KN06}; the SFE formulation has been extensively  studied with the goal of characterizing which functions can be securely evaluated \cite{Yao82,ben1988completeness,GMW87,Chaum88}.
One drawback of SFE is that depending on what auxiliary information the adversary might have, disclosing the exact function output might reveal 
each party's data. 
For example, consider computing the average of the data owned by all the parties. 
Even if we use SFE, a party's data can be recovered if all the other parties collaborate. 
To ensure protection of the private data under such a strong adversary, we want to impose a stronger privacy guarantee of differential privacy. 

Recent breaches of sensitive information about individuals due to linkage attacks prove the vulnerability of existing ad-hoc privatization schemes, such as anonymization of the records. 
In linkage attacks, an adversary matches up anonymized records containing sensitive information with public records in a different dataset. 
Such attacks have revealed the medical record of a former governor of Massachusetts \cite{Swe97}, the purchase history of Amazon users\cite{CKN11}, genomic information \cite{HSR08}, 
and movie viewing history of Netflix users \cite{NS08}.   
Differential privacy is a relatively recent formulation that has received considerable attention 
as a formal mathematical notion of privacy that provides protection against such strong adversaries (a recent survey is available at \cite{Dwo11}). 
The basic idea is to introduce enough randomness in the communication so that 
an adversary possessing arbitrary side information and access to the entire transcript of the communication will still have some residual uncertainty in identifying any of the bits at one of the parties.
 The requirement is strong enough that non-trivial functions will be computed 
only with some error. Thus, there is a great need for understanding the fundamental tradeoff between privacy
 and accuracy, and for the design of corresponding privatization mechanisms and communication protocols that achieve the optimal tradeoffs. 
This is the focus of this paper in the MPC context where each of the honest-but-curious parties possesses a {\em single bit} of information.

We study the following problem of multi-party computation under differential privacy: 
each party possesses a single bit of information; the information
bits are statistically independent. Each party is interested in computing a
function, which could differ from party to party, and there could be a
central observer (observing the entire transcript of the interactive
communication protocol)  interested in computing a separate function.
Performance at each party and the central observer is measured via the
accuracy of the function computed. We allow an arbitrary cost metric to
measure the distortion between the true and the computed function value. Each
party has a differential privacy constraint on its information bit (the
privacy level could be different from party to party) -- i.e., there remains
an uncertainty in any specific party's bit even to an adversary that has
access to the transcript of the interactions and all the other parties' bits.
The interactive communication is achieved via a broadcast channel that all
the parties and the central observer can hear (this modeling is without loss
of generality -- since the differential privacy constraint protects against
an adversary that can listen to the entire transcript, the communication
between any two parties might as well be revealed to all the others). 
It is useful to distinguish between two types of communication protocols: {\em interactive} and {\em non-interactive}. 
We say a communication protocol is non-interactive if a message broadcasted by one party does not depend on 
the messages broadcasted by any other parties. In contrast, interactive protocols allows the messages at any stage of the communication to
depend on all the previous messages.

 Our main result is the exact optimality of a simple non-interactive protocol in terms of maximizing accuracy for given privacy levels: each party randomizes (sufficiently) and publishes its own bit. In other words: \begin{quote}
{\em  non-interactive randomized response is exactly optimal}.
 \end{quote}
 Each party and the central observer then separately compute their respective decision functions to maximize the appropriate notion of their accuracy measure. The optimality is very general: it holds for all types of functions, heterogeneous privacy conditions on  the parties, all types of cost metrics, and both average and worst-case (over the inputs) measures of accuracy.
 Finally, the optimality result is {\em simultaneous}, in terms of maximizing accuracy at each of the parties and the central observer.
 Each party only needs to know its own desired level of privacy, its own function to be computed, and its measure of accuracy. Optimal data release and optimal decision making is naturally separated.
 
 The key technical result is a geometric understanding of the space of conditional probabilities of a given transcript: the interactive nature of the communication constrains the space to be a rank-1 tensor (a special case of Equation (6) in \cite{PP12} and perhaps implicitly used  in \cite{KQR09}; the two-party analog of this result is in \cite{Kil00}), while differential privacy imposes linear constraints on the singular vectors of this tensor. We characterize the convex hull of such manifolds of rank-1 tensors and show that their corner-points exactly correspond to the transcripts that arise from a non-interactive randomized response protocol.
 This  universal (for all functionalities) characterization is then used to argue that both average-case and worst-case accuracies are maximized by  non-interactive randomized responses. 

This geometric understanding leads to the novel linear program formulation of \eqref{eq:primal-lp} and \eqref{eq:primal-lp-wc}. 
Formulating utility maximization under differential privacy as linear programs has been previously studied in \cite{MT07,GRS09,BN10,GS10,GV13,GV12},  
under the standard client-server model where there is a single data publisher and a single data analyst. 
These approaches exploit the fact that 
both the differential privacy constraints and the utilities are linear in the matrix representing  
a privatization mechanism. 
Such a naive approach fails in this multi-party context, since multi-party protocols must satisfy additional non-linear constraints, namely 
the matrix describing the protocol must be compatible with multi-party (possibly interactive) communications. 
Mathematically, these constraints translate into certain rank constraints on higher order tensors, 
which are notoriously difficult to handle. 
The resulting maximization problem is non-linear and non-standard, i.e. the rank-1 constraints are not convex. 
Nevertheless, we introduce innovative linear program formulations of the problem, 
while paying the price in the increased complexity: the linear program is now infinite dimensional. 
Perhaps surprisingly, we prove that this infinite dimensional linear program 
has a simple optimal solution which we call {\em randomized response},  
by exploiting the geometric understanding of the manifold of rank-1 tensors. 
Upon receiving the randomized responses, each party can compute the best approximation of its respective function. 
A similar technique of transforming a non-linear optimization problem into an infinite dimensional LP has been 
successfully applied in \cite{KOV14}, where optimal privatization mechanisms under local differential privacy has been studied. 

Our main result shows that the optimal solutions of these infinite dimensional linear programs   
are at the corner points of the manifold of rank-1 tensors, which 
exactly correspond to the transcripts that arise from a non-interactive randomized response protocol. 
When the accuracy is measured via average accuracy, 
both the objective and the constraints are linear and 
it is natural to expect the optimal solution to 
be at the corner points (see Equation \eqref{eq:primal-lp}). 
A surprising aspect of our main result is that  
the optimal solution is still at the corner points 
even though the worst-case accuracy is a concave function over the protocol $P$ (see Equation \eqref{eq:nonconvex}). 

 This work focuses on the scenario where each party possesses a single bit of information. With multiple bits of information at each of the parties, the existence of a differentially private protocol with a fixed accuracy  
for any non-trivial functionality implies the existence of a protocol with the same level of privacy and same level of accuracy  for a specific functionality that only depends on one bit of each of the parties 
(as in \cite{GMPS}). 
Thus, if we can obtain lower bounds on accuracy for functionalities involving only a single bit at each of the parties, we obtain lower bounds on accuracy for all non-trivial general functionalities.  However, non-interactive communication is unlikely to be exactly optimal  in this general case where each party possesses multiple bits of information, and we provide a further discussion in Section \ref{sec:discussion}. 


\bigskip\noindent
{\bf Related Work.}  
In the context of two parties, privacy-accuracy tradeoffs  have been studied 
in \cite{MMPT,GMPS} where a single function is computed by a ``third-party" observing the transcript of the interactive protocol. \cite{MMPT} constructs natural functions that can only be computed very coarsely (using a natural notion of accuracy) as compared to a client-server model 
(which is essentially the single party setting).
 \cite{GMPS} shows that every non-trivial boolean functionality always incurs some loss of accuracy for any non-trivial privacy setting. Further, focusing on the specific scenario where each of the two parties has a single bit of information, \cite{GMPS} characterizes the {\em exact} accuracy-privacy tradeoff for AND and XOR functionalities; the corresponding optimal protocol turns out to be {\em non-interactive}. However, 
this result was derived under some assumptions: only two parties are involved, only the central observer computes an approximation of a function, the function has to be either XOR or AND, symmetric privacy conditions were used for both of the parties, and accuracy was measured only as worst-case over the four possible inputs.
Further, their analysis technique does not generalize to the case when we have more than two parties. 
To this end, we provide a new analysis technique of transforming the rank constrained optimization problem into a linear program, and give the exact optimal protocols for any number of parties, any function of interest, heterogeneous privacy requirements, and both average and worst-case accuracy measures. 
Among other things, this fully recovers the main results of \cite{GMPS} and does it with a more efficient protocol as discussed in Section \ref{sec:twoparty}.

While there is a vast literature on differential privacy in a variety of contexts, exact optimality results are very few. In an early result, \cite{GRS09} shows that adding discrete Laplacian noise to scalar 
count queries (which are a special case of integer functionalities with sensitivity one)  is universally optimal in terms of maximizing the average accuracy for any cost metric that is monotonic in the error. While such universal mechanisms do not exist in terms of maximizing average accuracy \cite{BN10}, recent work \cite{GV12,GV13} construct a class of mechanisms (termed as ``staircase" mechanisms)
 that are universally optimal in terms of maximizing worst-case accuracy for any cost metric that is monotonic in the error. Demonstrating a fundamental equivalence between binary hypothesis testing and differential privacy, 
\cite{OV13} derives data processing inequalities for differential privacy that are used to derive optimal composition theorems (characterization of how privacy degrades due to interactive querying). These techniques are also useful in the results derived in this paper.

The study of accuracy-privacy tradeoffs in the MPC context was first initiated by \cite{BNO08}  (addressed in a more general context earlier in \cite{DKMMN}) which studied a specific paradigm where differential privacy and SFE 
co-exist: the function to compute is decided from differentially private analyses and the method to compute it is decided from SFE theory. Specific functions such as SUM were studied in this setting, but no exact optimality results are available.
Exact optimality of non-interactive communication is demonstrated for two-party AND and XOR function computations in \cite{GMPS}. A curious fact in the context of AND computation is that  \cite{GMPS} requires the randomization of the bit to be in an output space of {\em three} letters (as opposed to the binary alphabet in standard randomized response).  At a first glance, this appears to be in contradiction to the claim in this paper. A closer look reveals that randomized response also achieves the same performance (worst-case accuracy over the four inputs) when combined with a different (and randomized) decision function. Indeed, the techniques from 
\cite{OV13} allow one to foresee this from an abstract point of view: every differentially private mechanism of a bit can be {\em simulated} from the output of randomized response with the same level of privacy. In other words, if $b$ is the bit, and $X$ is the (random) output of randomized response and $Y$ is the (random) output of some differentially private mechanism operating on $b$, then there exists a joint distribution on $(X,Y)$ such that the Markov chain $b-X-Y$ holds. This is discussed in detail in a later discussion section.

Function approximation has been widely studied in differential privacy literature 
under a centralized model where there is a single trusted entity owning a statistical database over a large number of individuals. Under this centralized 
setting, statistical learning has also been widely studied in differential privacy, e.g.  classification \cite{KLNRS11,CMS11}, k-means clustering \cite{BDMN05}, principal component analysis \cite{CSS12,CSS13,HR12,KT13}. 
In particular, it has been shown in \cite{KLNRS11} that under the centralized setting there exists a class of concepts that is efficiently learnable by {\em interactive} algorithms whereas 
a non-interactive algorithm requires exponential number of samples. 
An algorithm is called interactive in the centralized model,
 if it involves multiple rounds of communications between the server and the client. 
In contrast, we consider a multi-party setting where privacy barrier is on each individual owning his/her own data.
All communication happens in multiple rounds in multi-party computation, and a protocol is called interactive in the multi-party setting if one party's message 
depends on other party's previous messages. In this sense, the notion of interaction in multi-party computation is significantly different from what has been previously studied under centralized client-server settings. 

\section{Problem formulation}

Consider the setting where we have $k$ parties, each with its own private binary  data
$x_i\in\{0,1\}$ generated independently.
The independence assumption here is necessary
because without it each party can learn something about others,
which violates differential privacy, even without revealing any information. 
We discuss possible extensions to correlated sources in Section \ref{sec:discussion}.
Differential privacy implicitly imposes independence in a multi-party setting.
The goal of the private multi-party computation is for each party $i\in[k]$ to compute
an arbitrary function $f_i:\{0,1\}^k\to \cY$ of interest
by interactively broadcasting messages, while preserving the privacy of each party.
There might be a central observer who listens to all the messages being broadcasted,
and wants to compute another arbitrary function $f_0:\{0,1\}\to\cY$.
The $k$ parties  are honest in the sense that once they agree on what protocol to follow,
every party follows the rules.
At the same time, they can be curious, and
each party needs to ensure other parties cannot learn his bit
with sufficient confidence.
This privacy constraints are local differential privacy setting studied in \cite{DJW13} in the sense that 
there are multiple privacy barriers, each one separating 
each individual party and the rest of the world.
However, the main difference is that we consider multi-party computation, where 
there are multiple functions to be computed, and each node might possess a different function to be computed. 

Let $x=[x_1,\ldots,x_k]\in\{0,1\}^k$ denote the vector of $k$ bits,
and $x_{-i}=[x_1,\ldots,x_{i-1},x_{i+1},\ldots,x_k]\in\{0,1\}^{k-1}$ is the vector of bits except
for the $i$-th bit.
The parties agree on an interactive protocol to achieve the goal of multi-party computation.
A `transcript' is the output of the protocol, and
is a random instance of all broadcasted messages until all the communication terminates.
The probability that a transcript $\tau$ is broadcasted (via a series of interactive communications)
when the data is $x$
is denoted by
$P_{x,\tau} = \prob(\tau\,|\,x)$ for $x\in\{0,1\}^k$ and for $\tau\in \cT$.
Then, a protocol can be represented as a matrix denoting the probability distribution over
a set of transcripts $\cT$ conditioned on $x$: $P=[P_{x,\tau}] \in[0,1]^{2^k\times|\cT|}$.

In the end, each party makes a decision on what the value of function $f_i$
is, based on its own bit $x_i$ and the transcript $\tau$ that was
broadcasted. A decision rule is a mapping from a transcript $\tau\in\cT$ and
private bit $x_i\in\{0,1\}$ to a decision $y\in\cY$ represented by a function
$\hf_i(\tau,x_i)$. We allow randomized decision rules, in which case
$\hf_i(\tau,x_i)$ can be a random variable. For the central observer, a
decision rule is a function of just the transcript, denoted by a function
$\hf_0(\tau)$.

We consider two notions of accuracy: the average accuracy and the worst-case accuracy.
For the $i$-th party, consider an accuracy measure $\w_i:\cY\times\cY \to \reals$ (or equivalently a negative cost function) such that
$\w_i(f_i(x),\hf_i(\tau,x_i))$ measures the accuracy when the function to be computed is $f_i(x)$ and
the approximation is $\hf_i(\tau,x_i)$.
Then the average  accuracy for this $i$-th party is defined as
\begin{eqnarray}
	\aveacc(P,w_i,f_i,\hf_i) & \equiv & \frac{1}{2^k}\sum_{x\in\{0,1\}^k} \E_{\hf_i,P_{x,\tau} }[\w_i{(f_i(x),\hf_i(\tau,x_i))}] \;,	\label{eq:aveacc}
\end{eqnarray}
where the expectation is taken over the random transcript $\tau$ distribution as $P$ and also any randomness in the decision function $\hf_i$.
For example, if the accuracy measure is an indicator such that $w_i(y,y')=\ind_{(y=y')}$,
then $\aveacc$ measures the average probability of getting the correct function output.
For a given protocol $P$,
it takes $(2^k \,|\cT|)$ operations to compute the optimal decision rule:
\begin{eqnarray}
	f^*_{i,\rm ave}(\tau,x_i) &=& \arg\max_{y\in\cY} \sum_{x_{-i}\in\{0,1\}^{k-1}} P_{x,\tau} \,\w_i(f_i(x),y)\;,\label{eq:optave}
\end{eqnarray}
for each $i\in[k]$. The computational cost of $(2^k \,|\cT|)$ for computing
the optimal decision rule is unavoidable in general, since that is the
inherent complexity of the problem: describing the distribution of the
transcript requires the same cost. We will show that the optimal protocol
requires a set of transcripts of size $|\cT|=2^k$, and the computational
complexity of the decision rule for general a function is $2^{2k}$. However,
for a fixed protocol, this decision rule needs to be computed only once
before any message is transmitted. 
Further, it is also possible to find a closed form solution for the decision rule when 
$f$ has a simple structure. One example is the XOR function studied in detail in Section \ref{sec:XOR}, where 
the optimal decision rule is as simple as evaluating the XOR of all the received bits, which requires $O(k)$ operations. 
When there are multiple maximizers $y$, we
can choose arbitrarily, and it follows that there is no gain in randomizing
the decision rule for average accuracy. Similarly, the worst-case accuracy is
defined as
\begin{eqnarray}
	\wcacc(P,\w_i,f_i,\hf_i) & \equiv & \min_{x\in\{0,1\}^k} \E_{\hf_i,P_{x,\tau}}[\w_i{(f_i(x),\hf_i(\tau,x_i))}] \;.	\label{eq:wcacc}
\end{eqnarray}
For worst-case accuracy, given a protocol $P$, the optimal decision rule of the $i$-th party with a bit $x_i$  
can be computed by solving the following convex program:
\begin{eqnarray}
	Q^{(x_i)} = \underset{Q \,\in\, \reals^{|\cT|\times|\cY|}}{\text{arg max}} &&  
	\min_{x_{-i}\in\{0,1\}^{k-1}} \sum_{\tau\in\cT}\sum_{y\in\cY} P_{x,\tau} \,\w_i(f_i (x),y) Q_{\tau,y} 
	\label{eq:optwc} \\
	\text{subject to}&&  \sum_{y\in\cY} Q_{\tau,y}=1 \;,\;\forall \tau\in\cT \text{ and } Q\geq 0 \nonumber
\end{eqnarray}
The optimal (random) decision rule $f^*_{i,\rm wc}(\tau,x_i)$ is to output $y$ given transcript $\tau$ according to $\prob(y|\tau,x_i)=Q^{(x_i)}_{\tau,y}$. 
This can be formulated as a linear program with $(|\cT|\,|\cY|)$ variables and $(2^k+|\cT|)$ constraints. 
Again, it is possible to find a closed form solution for the decision rule when 
$f$ has a simple structure: for the XOR function, 
the optimal decision rule is again evaluating the XOR of all the received bits requiring $O(k)$ operations. 
For a central observer, the accuracy measures are defined similarly, and the optimal decision rule is now
\begin{eqnarray}
	f^*_{0,\rm ave}(\tau) &=& \arg\max_{y\in\cY} \sum_{x\in\{0,1\}^{k}} P_{x,\tau} \,\w_0(f_0(x),y)\;,   \label{eq:scenariotwo1}
\end{eqnarray}
and for worst-case accuracy the optimal (random) decision rule $f^*_{0,\rm wc}(\tau)$ is 
to output $y$ given transcript $\tau$ according to $\prob(y|\tau)=Q^{(0)}_{\tau,y}$.  
\begin{eqnarray}
	Q^{(0)} = \underset{Q \,\in\, \reals^{|\cT|\times|\cY|}}{\text{arg max}} &&  
	\min_{x\in\{0,1\}^{k}} \sum_{\tau\in\cT}\sum_{y\in\cY} P_{x,\tau} \,\w_0(f_0 (x),y) Q_{\tau,y} 
	\label{eq:scenariotwo2} \\
	\text{subject to}&&  \sum_{y\in\cY} Q_{\tau,y}=1 \;,\;\forall \tau\in\cT \text{ and } Q\geq 0 \nonumber
\end{eqnarray}
where $\w_0:\cY\times\cY\to\reals$ is the measure of accuracy for the central observer.

Privacy is measured by differential privacy \cite{Dwo06,DMNS06}.
Since we allow heterogeneous privacy constraints,
we use $\varepsilon_i$ to denote the desired privacy level of the $i$-th party.
We say a protocol $P$ is $\varepsilon_i$-differentially private for the $i$-th party if
for $i\in[k]$, and all  $x_i,x_i'\in\{0,1\}$, $x_{-i}\in\{0,1\}^{k-1}$, and $\tau\in\cT$,
\begin{eqnarray}
	\prob(\tau|x_i,x_{-i}) &\leq& e^{\varepsilon_i} \, \prob(\tau|x_i',x_{-i}) \;. \label{eq:defdp}
\end{eqnarray}
This condition ensures  no adversary can infer the private data $x_i$ with high enough confidence,
no matter what auxiliary information he might have and independent of his computational power.
To lighten notations, we let $\lambda_i=e^{\varepsilon_i}$ and say a protocol is $\lambda_i$-differentially private for the $i$-th party.
If the protocol is $\lambda_i$-differentially private for all $i\in[k]$, then
we say that the protocol is $\{\lambda_i\}$-differentially private for all parties.

A necessary condition on the multi-party protocols $P$, when the bits are generated independent of each other, is protocol compatibility \cite{GMPS}:
conditioned on the transcript of the protocol, the input bits stay independent of each other. Mathematically, a protocol $P$ is protocol compatible if each column $P^{(\tau)}$ is a {\em rank-one tensor},
when reshaped into a $k$-th order tensor $P^{(\tau)} \in [0,1]^{2\times2\times\ldots\times2}$,
where 
\begin{eqnarray}
	P^{(\tau)}_{x_1,\ldots,x_k} &=& P_{x,\tau}\;. \label{eq:tensor}
\end{eqnarray}
Precisely, there exist vectors $u^{(1)}\ldots, u^{(k)}$ such that $P^{(\tau)} = u^{(1)}\otimes \cdots \otimes u^{(k)}$,
where $\otimes$ denotes the standard outer-product, i.e. 
$P^{(\tau)}_{i_1,\ldots,i_k} = u^{(1)}_{i_1} \times \cdots\times u^{(k)}_{i_k}$.
This is crucial in deriving the main results, and  it is a well-known fact in the secure multi-party computation literature.
This follows from the fact that when the bits are generated independently,
all the bits are still independent conditioned on the transcript, i.e. $P(x|\tau)=\prod_i P( x_i |\tau)$,
which follows
implicitly from \cite{KQR09} and directly from Equation~(6) of \cite{PP12}.
For example, for a two-party case where $P(x|\tau) = P( x_1 |\tau) P(x_2 | \tau ) $,
\begin{eqnarray*}
	P =
	\begin{bmatrix}
	P(\tau|00) & P(\tau|01) \\
	P(\tau|10) & P(\tau|11)
	\end{bmatrix}
	=
	4P(\tau)
	\begin{bmatrix}
	P(00|\tau) & P(01|\tau) \\
	P(10|\tau) & P(11|\tau)
	\end{bmatrix}	
	=
	4P(\tau)
	\begin{bmatrix}
	P(x_1=0|\tau) \\
	P(x_1=1|\tau)
	\end{bmatrix}
	\begin{bmatrix}
	P(x_2=0|\tau) & P(x_2=1|\tau) \\
	\end{bmatrix}
\end{eqnarray*}
Notice that using the rank-one tensor representation of each column of the protocol $P^{(\tau)}$, we have $P(\tau|x_i=0,x_{-i})/P(\tau|x_i=1,x_{-i}) = u^{(i)}_1/u^{(i)}_2$.
It follows that $P$ is $\lambda_i$-differentially private if and only if
$ \lambda_i^{-1}  u^{(i)}_2 \leq u^{(i)}_ 1 \leq \lambda_i u^{(i)}_2$.

\bigskip
\noindent{\bf Randomized response.}
Consider the following simple protocol known as the {\em randomized response},
which is a term first coined by Warner \cite{War65} and commonly used in many private communications including the multi-party setting \cite{MMPT}. We will show in Section \ref{sec:main} that this is the optimal protocol for simultaneously maximizing the accuracy of all the parties. 
Each party broadcasts a randomized version of its bit
denoted by $\tx_i$ such that 
 \begin{eqnarray}
 	\tx_i = \left\{
	\begin{array}{rl}
		x_i & \text{ with probability }\frac{\lambda_i}{1+\lambda_i}\;,\\
		\bar{x_i} &\text{ with probability } \frac{1}{1+\lambda_i}\;,
	\end{array}
	\right.
	\label{eq:rr}
 \end{eqnarray}
 where $\bar{x_i}$ is the logical complement of $x_i$.
 Each transcript can be represented by the output of the protocol, which in this case is $\tx=[\tx_1,\ldots,\tx_k]\in\cT$,
 where $\cT=\{0,1\}^k$ is now the set of all broadcasted bits.
For example, in a simple case where $k=2$ and $\lambda_1=\lambda_2=\lambda$,
we have
\begin{eqnarray*}
	P &=& \frac{1}{(1+\lambda)^2} \begin{bmatrix}
		\lambda^2 	& \lambda		& \lambda 	& 1\\
		\lambda    		& \lambda^2 	& 1 			& \lambda\\
		\lambda 		& 1 			& \lambda^2 	& \lambda \\
		1 			&  \lambda 	& \lambda 	& \lambda^2 \\
	\end{bmatrix} \;\; \;,
\end{eqnarray*}
and the first column can be represented as a rank-one 2nd order tensor (which is a matrix) as
\begin{eqnarray*}
	P^{(00)} &=&
	\begin{bmatrix}
	\prob(\tx=00|x=00) & \prob(\tx=00|x=01)\\
	\prob(\tx=00|x=10)& \prob(\tx=00|x=11)\\
	\end{bmatrix}
	\;=\;\frac{1}{(1+\lambda)^2}
	\begin{bmatrix}
	 \lambda^2 & \lambda \\
	 \lambda & 1\\
	\end{bmatrix} \;=\;
	\frac{1}{(1+\lambda)^2}
	\begin{bmatrix}
	\lambda\\ 1\\
	\end{bmatrix}
	\begin{bmatrix}
	\lambda & 1
	\end{bmatrix}\;.
\end{eqnarray*}
This confirms that the first column of $P$ is a rank-one matrix $P^{(00)}$ with $u^{(1)}=(1/(1+\lambda))[\lambda\;,\; 1]$ and
$u^{(2)}=(1/(1+\lambda)) [\lambda\;,\; 1]$, hence protocol compatible.
Also notice that it satisfies the differential privacy constraints, since $\lambda^{-1} u^{(i)}_2\leq u^{(i)}_1 \leq  \lambda u^{(i)}_2$.

\bigskip\noindent
{\bf Accuracy maximization.}
Consider the  problem of maximizing the average accuracy for a centralized observer with function $f$. 
Up to the scaling of $1/2^k$ in \eqref{eq:aveacc}, the accuracy can be written as 
\begin{eqnarray}
 \label{eq:objective}
	\sum_{x\in\{0,1\}^k} \E_P[w{(f(x),\hf_0(\tau))}]
	&=& \sum_{x}\sum_{y\in\cY}
	\underbrace{w{(f_0(x),y)}}_{\triangleq \, W_{x}^{(y)}} \sum_{\tau\in\cT}
	P_{x,\tau} \underbrace{\prob(\hf_0(\tau)=y)}_{\triangleq \,Q_{\tau,y}} \;,\; 
\end{eqnarray}
where $\hf_0(\tau)$ denotes the randomized decision up on receiving the transcript $\tau$. 
In the following we define 
$W_x^{(y)} \triangleq w(f_0(x),y)$ to represent the accuracy measure 
and $Q_{\tau,y}\triangleq \prob(\hf(\tau)=y)$ to represent the decision rule. 

Focusing on this single central observer for the purpose of illustration, 
we want to design protocols $P_{x,\tau}$ and decision rules $Q_{\tau,y}$ that maximize the above accuracy. 
Further, this protocol has to be compatible with interactive communication, 
satisfying the rank one condition discussed above, 
and satisfy the differential privacy condition in \eqref{eq:defdp}.
Hence, we can formulate the accuracy maximization can be formulated as follows 
given $W_x^{(y)}$'s in terms of 
the function $f_0(\cdot)$ to be computed and   
an accuracy measure $w_0(\cdot,\cdot)$, 
and required privacy level $\lambda_i$'s:
\begin{equation}
\begin{aligned}
& \underset{ P \in \reals^{2^k\times|\cT|}, Q \in\reals^{|\cT|\times|\cY|} }{\text{maximize}}
& & \sum_{x,\in\{0,1\}^k, y\in\cY} W^{(y)}_x\; \sum_{\tau\in\cT} P_{x,\tau}Q_{\tau,y} \\
& \text{subject to}
& & P \text{ and } Q \text{ are row-stochastic matrices}\;, \\
& & & \text{rank}(P^{(\tau)})=1 \;,\;  \forall \tau\in\cT \;, \\
& & & P_{(x_i,x_{-i}),\tau} \leq {\lambda_i} P_{(x_i',x_{-i}),\tau}\;, \;\; \forall i\in[k], x_1,x_1',\in\{0,1\},x_{-i}\in\{0,1\}^{k-1} \text{ and } \tau\in\cT\;,
\end{aligned}
\label{eq:optimization}
\end{equation}
where $P^{(\tau)}$ is defined as a $k$-th order tensor defined from the $\tau$-th column of matrix $P$ as defined in Equation \eqref{eq:tensor}. 
Notice that the rank constraint is only a necessary condition for a protocol to be compatible with interactive communication schemes, 
i.e. a valid interactive communication protocol implies the rank-one condition but 
not all rank-one protocols are valid interactive communication schemes.  
Therefore, the above is a relaxation with larger feasible set of protocols, 
but in turns out that the optimal solution of the above optimization problem 
is the randomized response, which is a valid (non-interactive) communication protocol. 
Hence, there is no loss in solving the above relaxation. 

The main challenge in solving this optimization is that 
it is a rank-constrained tensor optimization which is notoriously difficult. 
Since the rank constraint is over a $k$-th order tensor ($k$-dimensional array) with possibly $k>2$, 
common approaches of convex relaxation from \cite{RFP10} for matrices (which are 2nd order tensors) 
does not apply.  
Further, we want to simultaneously apply similar optimizations 
to all the parties with different functions to be computed. 

We introduce a novel transformation of the above rank-constrained optimization 
into a linear program in  \eqref{eq:primal-lp} and \eqref{eq:primal-lp-wc}. 
The price we pay is in the increased dimensionality: the LP has an infinite dimensional decision variable. 
However, combined with the geometric understanding of the 
the manifold of rank-1 tensors, we can identify the exact optimal solution. 
We show in the next section that given desired level of privacy $\{\lambda_i\}_{i\in[k]}$,
 there is a single universal protocol that simultaneously 
maximizes the accuracy for $(a)$ all parties; $(b)$ any functions of interest; $(c)$ any accuracy measures; and  
$(d)$ both worst-case and average case accuracy. Together with optimal decision rules performed at each of the receiving ends, this gives the exact optimal multi-party computation scheme.

\section{Main Result}
\label{sec:main}

We show, perhaps surprisingly, that the simple randomized response presented in \eqref{eq:rr} is
the unique optimal protocol in a very general sense.
For any desired privacy level $\lambda_i$, and arbitrary function $f_i$, for any accuracy measure $\w_i$, and any notion of accuracy (either average or worst case), we show that the randomized response is universally optimal.

\begin{theorem}
	Let  the optimal decision rule be defined as in \eqref{eq:optave} for the average accuracy and
	\eqref{eq:optwc} for the worst-case accuracy.
	Then, for any $\lambda_i \geq 1$, any function $f_i:\{0,1\}^k\to \cY$, and any accuracy measure $\w_i:\cY\times\cY\to\reals$ 
	for $i\in[k]$,
	the randomized response for given $\lambda_i$ 
	with the optimal decision function
	achieves the maximum accuracy for the $i$-th party
	among all $\{\lambda_i\}$-differentially private interactive protocols  and all decision rules.
	For the central observer, the randomized response with the optimal decision rule defined as in \eqref{eq:scenariotwo1} and 
	\eqref{eq:scenariotwo2}
	achieves the maximum accuracy among
	all $\{\lambda_i\}$-differentially private interactive protocols  and all decision rules for any arbitrary function $f_0$ and any measure of accuracy $\w_0$.
	\label{thm:average}
\end{theorem}

This is a strong universal optimality. Every party and the central observer
can {\em simultaneously} achieve the optimal accuracy, using a universal
randomized response. Each party only needs to know its own desired level of
privacy, its own function to be computed, and its measure of accuracy.
Optimal data release and optimal decision making is naturally separated.
However, it is not immediate at all  that a non-interactive scheme such as the randomized response would
achieve the maximum accuracy.
We need to utilize the convex geometry of the problem, in order to show that interaction is not necessary.

Once we know that interaction does not help,
we can make an educated guess that the randomized response should dominate over other non-interactive schemes.
This intuition follows from the dominance of randomized response in the single-party setting,
that was proved using a
powerful operational interpretation of differential privacy first introduced in \cite{OV13}.
This intuition can in fact be made rigorous,
as we show in the following section with a simple two-party example.

\subsection{Proof of Theorem \ref{thm:average}} 

We first focus on the 
scenario where a central observer wants to compute a function $f$ over $k$ bits distributed across $k$ parties.
We will show in Section \ref{sec:averageproof} that $\aveacc(P,\w,f,\hf)$ is maximized when randomized response protocol is used with the optimal decision rule of
\eqref{eq:scenariotwo1}.
Subsequently in Section \ref{sec:worstproof}, we show that $\wcacc(P,\w,f,\hf)$ is maximized when again
randomized response protocol is used with the optimal decision rule of
\eqref{eq:scenariotwo2}.
Theorem \ref{thm:average} directly follows from these two results,
since the $i$-th party can
compute the optimal decision and achieve the maximum accuracy for each instance of $x_i\in\{0,1\}$.

\subsubsection{Proof for the average case}
\label{sec:averageproof}
\begin{theorem}
	For a central observer who wants to compute $f$ with accuracy measure $\w$,
	randomized response with  the optimal decision rule of \eqref{eq:scenariotwo1}
	maximizes the average accuracy $\aveacc(P,\w,f,\hf)$ among all $\{\lambda_i\}$-differentially private protocols and all decision rules.
\end{theorem}
In this section, we provide a proof of this theorem. 
We want to solve the rank-constrained optimization problem of \eqref{eq:optimization}.
The sketch of the proof is as follows. 
First, we introduce a novel change of variables to transform the optimization into an infinite dimensional linear program. 
Next, we show that if the optimal solution to this LP has non-zero probability only for `extremal' transcripts (see Definition \ref{def:extremal}), 
then 
there is only one possible protocol which is the randomized response in \eqref{eq:rr}. 
Finally, we finish the proof by using dual LP to prove that the optimal solution can only have non-zero probability at the 
`extremal' transcripts. 

\bigskip\noindent
{\bf LP formulation.}
We want to maximize the average accuracy over $P$ and $Q$, where the average accuracy is (up to a scaling by $1/2^k$) 
\begin{eqnarray*}
	\sum_{x} \E_P[w{(f(x),\hf(\tau))}]
	&=& \sum_{x}\sum_{y\in\cY}
	\underbrace{w{(f(x),y)}}_{\triangleq\,W_{x}^{(y)}} \sum_{\tau\in\cT}
	P_{x,\tau} \underbrace{\prob(\hf(\tau)=y)}_{\triangleq\,Q_{\tau,y}} \;\; = \; \sum_{y} \Big\langle W^{(y)} , \sum_\tau P_\tau Q_{\tau,y} \Big\rangle\;,
\end{eqnarray*}
	where $\langle ,\rangle$ denote the standard inner product such that 
	$\Big\langle W^{(y)} ,  P_\tau Q_{\tau,y} \Big\rangle = \sum_x \big( W^{(y)}_x   P_{x,\tau} Q_{\tau,y} \big)$, and $P_\tau$ is the column of the matrix $P$ corresponding to $\tau$.
	The $2^k\times|\cT|$-dimensional matrix $P$ represents the 
	conditional distribution of the transcripts $\tau$ given the original data $x$, such that 
	$P_{x,\tau} = \prob(\tau|x)$. 
	The $|\cT|\times |\cY|$-dimensional matrix $Q$ represents the decision rule, possibly randomized.
	For example, if we consider two-party XOR computation with the same level of privacy $\lambda$, 
	a solution (which turns out to be optimal) is randomized response with decision rule according to the XOR of the received bits. 
	In particular, $\tau\in\{00,01,10,11\}$ and $y=\hf(\tau)$ is the XOR of the two bits in $\tau$. 
	This can be written as 
	\begin{eqnarray}
		P \;=\; \frac{1}{(1+\lambda)^2}\begin{bmatrix} 
			\lambda^2 & \lambda & \lambda & 1\\
			\lambda & \lambda^2 & 1 & \lambda \\
			\lambda & 1 & \lambda^2 & \lambda \\
			1 & \lambda & \lambda & \lambda^2
		\end{bmatrix}
		\;,\;\text{ and } \;
		Q \;=\;  \begin{bmatrix} 
			1&0 \\
			0& 1\\
			0& 1\\
			1& 0
		\end{bmatrix}\;.
		\label{eq:XOR2}
	\end{eqnarray}
	Notice that the labeling of 
	$\tau$ is arbitrary and applying the same permutation to the columns of $P$ and the rows of $Q$ does not change the feasibility 
	or the accuracy of the solution. 
	The columns of $P$ are still rank one when written in an appropriate tensor form, and also satisfy the differential privacy 
	constraints. 
	Another important point is that we cannot restrict the number of transcripts a priori, and when solving \eqref{eq:optimization}, we 
	need to consider infinite dimensional (but countable) $\cT=\Z$. 
	The objective and the constraints depend on 	
	\begin{eqnarray*}
		 [\prob(y,\hf(\tau)=y|x)]_{x,\tau,y} \;=\; [P_{x,\tau}Q_{\tau,y}]_{x,\tau,y} \;,
	\end{eqnarray*}
	for	$x\in\{0,1\}^k$, $\tau\in\Z$, and $y\in\cY$ 
	where how we label or index the transcript $\tau$ is arbitrary. 
	Since the rank constraints on the tensorized version of the columns of $P$ are difficult to handle, 
	we exploit the fact that the problem is invariant in renaming of the transcript index $\tau$, 
	and introduce a new indexing of the transcripts and new representation of the 
	effective decision variable $[\prob(y,\hf(\tau)=y|x)]_{x,\tau,y}$. 
	
	Define a {\em signature vector} as a vector $S_{(s_1,\ldots,s_k)} \in \reals^{2^k}$ indexed by 
	${(s_1,\ldots,s_k)}\in [\lambda_1^{-1},\lambda_1]\times\cdots\times[\lambda_k^{-1},\lambda_k]$. 
  	A signature vector $S_{s_1,\ldots,s_k}$ is a vectorized version of a rank-one tensor $[1\,,\,s_1]\otimes \cdots \otimes [1\,,\, s_k]$
	(to ensure that the rank constraint is satisfied) 
	with $\lambda_i^{-1}\leq s_i \leq \lambda_i$ for all $i\in[k]$ (to ensure that the differential privacy constraint is satisfied). 
	The index $(s_1,\ldots,s_k)$ effectively  replaces the indexing of the transcript $\tau$.  
		Consider an infinite dimensional  matrix $S$, where the number of rows is $2^k$ and 
	the number of columns is uncountably infinite.  
	The {\em signature matrix} $S$ contains as its columns all possible choices of the signature vector 
	$S_{(s_1,\ldots,s_k)}$ indexed by $(s_1,\ldots,s_k)$. 
	Given this definition  $S$, 
	the space of all possible feasible protocols and all possible corresponding decision rules 
	can be represented as 
	\begin{eqnarray}
		 [\prob(y,\hf(\tau)=y|x)]_{x,\tau,y} \;=\; [S_{x,(s_1,\ldots,s_k)}\theta^{(y)}_{(s_1,\ldots,s_k)}]_{x,(s_1,\ldots,s_k),y} \;,
		 \label{eq:connect}
	\end{eqnarray}
	where the equality is up to a appropriate mapping of indexes in $\tau$ and $(s_1,\ldots,s_k)$  
	and merging/splitting/dropping of appropriate columns. 
	As a concrete example, the conditional distribution of outputting $y=0$ in \eqref{eq:XOR2} is 
	\begin{eqnarray} 
		 [\prob(\tau,\hf(\tau)=y|x)]_{x,\tau,y=0} \;=\; P\diag(Q^{(0)}) \;=\; 
		 \frac{1}{(1+\lambda)^2}\begin{bmatrix} 
			\lambda^2 & 0 & 0 & 1\\
			\lambda & 0 & 0 & \lambda \\
			\lambda & 0 & 0 & \lambda \\
			1 & 0 & 0 & \lambda^2
		\end{bmatrix}\;,
		\label{eq:XOR3}
	\end{eqnarray}
	which can be represented (up to a reindexing of the columns) using the signature matrix as 
	\begin{eqnarray}
		S \diag(\theta^{(0)}) \;=\; \begin{bmatrix} 
		1 & \cdots  & 1 &\cdots\\
		 \lambda^{-1} & \cdots & \lambda &\cdots \\
		 \lambda^{-1} & \cdots & \lambda &\cdots \\
		 \lambda^{-2} & \cdots & \lambda^2 &\cdots
		\end{bmatrix}
		\begin{bmatrix}
		\frac{\lambda^2}{(1+\lambda)^2} & &  & & \\
		& 0 \\
		& & \ddots \\
		& & & 0\\
		& & & & \frac{1}{(1+\lambda)^2} \\	
		& & & & & 0\\
		& & & & & & \ddots  \\
		\end{bmatrix}
	\end{eqnarray}
	For all practical purposes, these two matrices represent the same protocol and the same decision rule. 
	Since $S$ is a fixed matrix for given problem parameters $k$ and $\lambda_i$'s, 
	the new decision variable is just the set of scaling vectors $\{\theta^{(y)}\}_{y\in\cY}$. 
	By optimizing over $\theta^{(y)}$'s, we are effectively selecting a subset of signatures to include in our transcript, 
	and choosing the randomized outputs of those selected transcripts. 
	We want to maximize the average accuracy, conditioned on the fact that
	conditional probabilities sum to one   and probabilities are non-negative.
	\begin{equation}
	\begin{aligned}
	& \underset{\bb^{(1)},\ldots,\bb^{(|\cY|)}}{\text{maximize}}
	& & \sum_{y\in\cY} \la W^{(y)},S\theta^{(y)} \ra 
			\;=\; 
			\sum_{y\in\cY, (s_1,\ldots,s_k)}  (S^TW^{(y)})_{(s_1,\ldots,s_k)} \theta_{(s_1,\ldots,s_k)}^{(y)}  \\
	& \text{subject to}
	& &  \sum_{y\in\cY} \sum_{(s_1,\ldots,s_k)}\s_{(s_1,\ldots,s_k)} \bb^{(y)}_{(s_1,\ldots,s_k)} = \unity\\
	& & &  \bb^{(y)} \geq 0.
	\end{aligned}
	\label{eq:primal-lp}
	\end{equation}
	This is a linear program in $\theta^{(y)}$'s and once we have the optimal solution we can translate it to 
	the original variables using \eqref{eq:connect}. 
	However, numerically solving the above problem is infeasible since 
	the dimension of each variable $\theta^{(y)}$ is now uncountably infinite. 
	We first claim that the solution of this problem is simple and can be represented in a closed form, 
	and then prove this claim using the dual LP.

\begin{definition}
A $2^k$-dimensional column vector $S_{(s_1,\ldots,s_k)}$ is {\em extremal} if the $k$-th order 
tensorization of $S_{(s_1,\ldots,s_k)}$ is a rank-one tensor of the form $[1\;,\;s_1] \otimes \cdots \otimes [1\;,\; s_k] $ with factors
$s_i \in \{\lambda_i^{-1},\lambda_i\}$ for all $i\in[k]$. There are $2^k$ such extremal columns of $S$.
\label{def:extremal}
\end{definition}
This notion of extremal transcript is consistent with a similar notion of extremal privatization mechanisms defined 
in \cite{KOV14} as a set of mechanisms whose conditional distributions are at the extreme points of differential privacy constraints. 
When $k=2$ there are four extremal columns of $S$: 
\begin{eqnarray*}
	\begin{bmatrix}
		1\\
		\lambda_1\\
		\lambda_2\\
		\lambda_1\lambda_2\\
	\end{bmatrix}\;,\;
	\begin{bmatrix}
		1\\
		\lambda_1\\
		\lambda_2^{-1}\\
		\lambda_1\lambda_2^{-1}\\
	\end{bmatrix}\;,\;
	\begin{bmatrix}
		1\\
		\lambda_1^{-1}\\
		\lambda_2\\
		\lambda_1^{-1}\lambda_2\\
	\end{bmatrix}\;,\;
	\begin{bmatrix}
		1\\
		\lambda_1^{-1}\\
		\lambda_2^{-1}\\
		\lambda_1^{-1}\lambda_2^{-1}\\
	\end{bmatrix}.
\end{eqnarray*}
We make the following claim. 
\begin{remark}
	\label{rem:extremal}
	The optimal solution to the LP in \eqref{eq:primal-lp} only has strictly positive $\theta^{(y)}_{(s_1,\ldots,s_k)}$ 
	for $(s_1,\ldots,s_k)$ corresponding to 
	extremal columns of $S$ and all the non-extremal columns are set to zero. 
\end{remark}
Suppose for now that  this claim is true, then we can make following observations.
\begin{itemize}

 \item There is an optimal solution of the LP that requires no randomized decision.
 Suppose the set $\{\theta^{(y)}\}_{y\in\cY}$ is an optimal solution, and there is an extremal transcript $(s_1,\ldots,s_k)$ such that
 both $\theta_{(s_1,\ldots,s_k)}^{(y_1)}$ and $\theta_{(s_1,\ldots,s_k)}^{(y_2)}$ are non-zero 
 for some $y_1,y_2\in\cY$. Then, we can construct a new optimal solution
 by setting $\tilde{\theta}_{(s_1,\ldots,s_k)}^{(y_1)} = \theta_{(s_1,\ldots,s_k)}^{(y_1)} +\theta_{(s_1,\ldots,s_k)}^{(y_2)}$ and $\tilde{\theta}_{(s_1,\ldots,s_k)}^{(y_2)}=0$.
 Continuing in this fashion, we can construct an optimal solution with no randomization.

 \item Since the $2^k\times 2^k$ sub matrix of $S$ corresponding to the extremal columns is now an invertible matrix, 
 $\theta=\sum_{y\in\cY}\theta^{(y)}$ is easily computed by the equality constraint.
 Once the optimal $\theta$ is fixed,
 we can identify the optimal decision rule for each transcript separately.
 Among $\theta^{(y)}_{(s_1,\ldots,s_k)}$'s for $y\in\cY$, put all the mass on the $y$ that maximizes $(S^TW^{(y)})_{(s_1,\ldots,s_k)}$.
 The optimal protocol $S\diag(\theta)$ is uniquely determined, and
 finding the optimal decision rule (i.e. $\theta^{(y)}$) is also simple once we have the protocol.
 This gives the precise optimal decision rule described in Equation \eqref{eq:optave}.

 \item This uniquely determined optimal protocol is the randomized response defined in Equation \eqref{eq:rr} for 
 all possible choices of the problem parameters, and it is a non-interactive protocol.

\end{itemize}

\bigskip\noindent
{\bf Proof of the remark \ref{rem:extremal} using the geometry of the manifold of rank one tensors.}
Now, we are left to prove the claim that the optimal solution only contains the extremal signatures.
Consider a $k$-dimensional manifold in $2^k$-dimensional space:
\begin{eqnarray*}
\cM_{\{\lambda_i\}}  &=&  \{T: \text{$T=[1, t_1]\otimes\cdots\otimes[1,t_k]$ and $\lambda_i^{-1}\leq t_i\leq\lambda_i$ for all $i\in[k]$}\}\;, \\
\cP_{\{\lambda_i\}} &=& \text{conv}(\cM_{\{\lambda_i\}})\;,
\end{eqnarray*}
where $\text{conv}(\cdot)$ is the convex hull of a set. The following result characterizes the polytope $\cP_{\{\lambda_i\}}$, the proof of which is moved to Section~\ref{sec:polytope}.
\begin{lemma}
	\label{lem:polytope}
	The convex hull $\cP_{\{\lambda_i\}}$ is a polytope with $2^k$ faces and $2^k$ corner points corresponding to the $2^k$ extremal
	columns of $S$. Further, the intersection of the manifold $\cM_{\{\lambda_i\}}$ and the boundary of $\cP_{\{\lambda_i\}}$ 
	 is only the set of those corner points. Hence,
	any point in the manifold is represented as a convex combination of the corner points, 
	and it requires all the corner points to represent any point in the manifold that is not already one of the corner points. 
\end{lemma}
This implies that any column of $S$ can be represented as a convex combination of the extremal columns of $S$.
We can write the dual of the primal LP in Equation~\eqref{eq:primal-lp} as:
\begin{equation}
\begin{aligned}
\label{eq:dual1}
& \underset{\mu\in\reals^{2^k}}{\text{minimize}}
& & \sum_{x \in\{0,1\}^k}\mu_{x}   \\
& \text{subject to}
& & \la S_{(s_1,\ldots,s_k)} , \mu \ra \;\geq\;  \la S_{{(s_1,\ldots,s_k)}} \,,\,W^{(y)} \ra \;,\;\;\; \text{ for all }{y\in\cY,{(s_1,\ldots,s_k)}\in[\lambda_1^{-1},\lambda_1]\times\cdots\times[\lambda_k^{-1},\lambda_k]}.
\end{aligned}
\end{equation}
Consider an optimal dual solution $\mu^*$. 
We now prove that for any dual optimal solution, the constraints in Equation \eqref{eq:dual1}
can be met with equality only for the indices $(s_1,\ldots,s_k)$ corresponding to corner points of $\cP_{\{\lambda_i\}}$.
By complementary slackness of LP, this implies that
the primal variable $\theta^{(y)}_{(s_1,\ldots,s_k)}$ can only be strictly positive for the extremal transcripts,
and all non-extremal transcripts must be zero.

If $\la T , \mu^* \ra \;=\;  \sum_{x} W_{x}^{(y)} T_{x}$ for some $T\in\cM_{\{\lambda_i\}}$ which is not an extremal point,  
then it follows from Lemma \ref{lem:polytope} that $T$ can be represented as a convex combination of the extremal points. 
Unless all the constraints for $\mu^*$ are satisfied with equalities (which can only happen if $W^{(y)}$ are all same for all $y\in\cY$ 
and all protocols and  decision rules achieve the same accuracy), 
there exists at least one extremal signature $S_{(s_1,\ldots,s_k)}$ such that 
the inequality in \eqref{eq:dual1} is violated. Hence, it contradicts 
the assumption that $\mu^*$ is a feasible dual solution.


\subsubsection{Proof for the worst-case accuracy}
\label{sec:worstproof}
\begin{theorem}
	For a central observer who wants to compute $f$ with accuracy measure $\w$,
	randomized response with  the optimal decision rule of \eqref{eq:scenariotwo2}
	maximizes the worst-case accuracy $\wcacc(P,\w,f,\hf)$ among all $\{\lambda_i\}$-differentially private protocols and all decision rules.
\end{theorem}
In this section, we provide a proof of this theorem. 
Consider the worst case accuracy of the form
\begin{eqnarray*}
	\min_{x\in\{0,1\}^k} \E_{\hf(\tau)}[w{(f(x),\hf(\tau))}]
	&=& \min_{x}\sum_{y\in\cY}
	\underbrace{w{(f(x),y)}}_{W_{x}^{(y)}} \sum_{\tau\in\cT}
	P_{x,\tau} \underbrace{\prob(\hf(\tau)=y)}_{Q_{\tau,y}}\;.
\end{eqnarray*}
Using the signature matrix $S$, 
we can write this as maximizing a concave function (minimum over a set of linear functions is a concave function):
\begin{equation}
\begin{aligned}
& \underset{\bb^{(1)},\ldots,\bb^{(|\cY|)}}{\text{maximize}}
& & \min_{x\in\{0,1\}^k} \Big\{ \sum_{y\in\cY}  W^{(y)}_{x}\, \big(S\theta^{(y)}\big)_{x} \Big\}   \\
& \text{subject to}
& & \s \sum_{y\in\cY} \bb^{(y)} = \unity\\
& & &  \bb^{(y)} \geq 0.
\end{aligned}
\label{eq:nonconvex}
\end{equation}
This can be formulated as the following primal LP:
\begin{equation}
\begin{aligned}
\label{eq:primal-lp-wc}
& \underset{\xi,\bb^{(1)},\ldots,\bb^{(|\cY|)}}{\text{maximize}}
& & \xi   \\
& \text{subject to}
& & \xi \;\leq\;  \Big\{ \sum_{y\in\cY}  W^{(y)}_{x}\, \big(S\theta^{(y)}\big)_{x} \Big\} \;,\;\;\; \text{ for all }{x\in\{0,1\}^k}\\
& & & \s \sum_{y\in\cY} \bb^{(y)} = \unity\\
& & &  \bb^{(y)} \geq 0.
\end{aligned}
\end{equation}
Define dual variables $\nu\in\reals^{2^k}$ corresponding to the first set of constraints and $\mu\in\reals^{2^k}$ to the second. Then the dual LP is
\begin{equation}
\begin{aligned}
\label{eq:dual}
& \underset{\nu,\mu}{\text{minimize}}
& & \sum_{x}\mu_{x}  \\
& \text{subject to}
& & \la S_{(s_1,\ldots,s_k)} , \mu \ra \;\geq\;  \sum_{x} W_{x}^{(y)} S_{x ,{(s_1,\ldots,s_k)}} \nu_{x} \;,\;\;\; \text{ for all }{y\in\cY,{(s_1,\ldots,s_k)}\in[\lambda_1^{-1},\lambda_1]\times\cdots\times[\lambda_k^{-1},\lambda_k]}\\
& & & \unity^T \nu = 1\\
& & &  \nu \geq 0.
\end{aligned}
\end{equation}

Consider an optimal solution $(\nu^*,\mu^*)$. This defines a polytope for the each column of $S$ put in a tensor form in $\reals^{2^k}$:
$$ \cP_{\nu^*,\mu^*} = \{T: \la T , \mu^* \ra \;\geq\;  \sum_{x} W_{x}^{(y)} T_{x} \nu^*_{x},\text{ for all } y\in\cY \}.$$
Now, $(\nu^*,\mu^*)$ is feasible if and only if $\cP_{\{\lambda_i\}} \subseteq \cP_{\nu^*,\mu^*}$, since the condition must be met by all $\lambda$-DP protocol-compatible transcripts.

Since both $\cP_{\nu^*,\mu^*}$ and $\cP_{\{\lambda_i\}}$ are convex polytopes,
and $\cP_{\{\lambda_i\}} \subseteq \cP_{\nu^*,\mu^*}$ for our choice of optimal solutions,
then the constraints in
Eq. \eqref{eq:dual} can only be met with equality for signatures corresponding to the
intersection of $\cM_{\{\lambda_i\}}$ and the boundary of $\cP_{\nu^*,\mu^*}$. 
From Lemma \ref{lem:polytope}, we know that such intersection can only happen at the extremal points. 
By complementary slackness of LP, this implies that
the primal variable $\theta^{(y)}_{(s_1,\ldots,s_k)}$ can only be strictly positive for the extremal transcripts,
and all non-extremal transcripts must have zero value.
However, in this case, one might need to resort to randomized decisions depending on the accuracy weights $W$.

The optimality of the extremal protocols can 
also be also explained perhaps more intuitively as follows. 
Consider the primal LP formulation.
Let $\Theta=\{\theta^{(1)},\ldots,\theta^{(|\cY|)}\}$ be an optimal solution that has at least one value that is non-extremal.
Without loss of generality, let $\theta^{(1)}_i$ be the positive value corresponding to a non-extremal transcript $S_i$.
Then, by the lemma, we know that we can
represent $S_i=\sum_{j=1}^{2^k} \alpha_j S_j$, where $S_1\ldots,S_{2^k}$ are the extremal transcripts.
Then, we can construct another feasible solution $\tilde\Theta=\{\tilde\theta^{(1)},\ldots,\tilde\theta^{(|\cY|)}\}$
from $\Theta$, by taking the value of $\theta^{(1)}_i$ and add it to the extremal ones according to
$\tilde\theta^{(1)}_j = \theta^{(1)}_j + \alpha_j\theta^{(1)}_i$, and setting $\tilde\theta^{(1)}_i=0$.
The new solution preserves the summation $S\tilde\theta^{(y)} = S\theta^{(y)}$.
Since the new solution has one less non-extremal value, we can continue in this fashion until we are left with only extremal
transcripts.

\subsection{Two-party function computation}
\label{sec:twoparty}

In this section, we show that
randomized response always dominates over any other non-interactive schemes.
Precisely, we will show the following claim:
{\em for any non-interactive protocol and a decision rule,
there exists a randomized response and a decision rule for the randomized response
 that achieves the same accuracy,
for any privacy level, any function, and any measure of accuracy.}

The statement is generally true, but for concreteness we focus on a specific
example in the two-party setting, which captures all the main ideas. In this
setting, there are essentially only two functions of interest, AND and XOR,
and it is only interesting to consider the scenario where the central
observer is trying to compute these functions over two bits distributed
across two parties. Private AND function computation under the worst-case
accuracy measure was studied in \cite{GMPS}. The authors of \cite{GMPS}
proposed a non-interactive scheme and showed that it achieves the optimal
accuracy of $\lambda(\lambda^2+\lambda+2)/(1+\lambda)^3$ when both parties
satisfy $\lambda$-differential privacy.

We will show by example how to construct a randomized response that dominates
any non-interactive scheme. The protocol proposed in \cite{GMPS} outputs a
privatized version of each bit according to the following rule
\begin{eqnarray*}
	M(0) &=& \left\{
	\begin{array}{rl}
		0& \text{w.p.} \frac{\lambda}{1+\lambda}\\
		1& \text{w.p.} \frac{\lambda}{(1+\lambda)^2}\\
		2& \text{w.p.} \frac{1}{(1+\lambda)^2}
	\end{array}
	\right. \;\;,\text{ and }\;\;\;\;
	M(1) \;=\; \left\{
	\begin{array}{rl}
		0& \text{w.p.} \frac{1}{1+\lambda}\\
		1& \text{w.p.} \frac{\lambda^2}{(1+\lambda)^2}\\
		2& \text{w.p.} \frac{\lambda}{(1+\lambda)^2}
	\end{array}
	\right. \;,
\end{eqnarray*}
which satisfies $\lambda$-differential privacy. Such a non-interactive
protocol of revealing the privatized data is referred to as a privacy
mechanism. Upon receiving this data, the central observer makes a decision
according to
\begin{eqnarray*}
	\hf(M(x_1),M(x_2)) &=& \left\{
	\begin{array}{rl}
	1&\text{ if } M(x_1)=2 \text{ or } M(x_2)=2\\
	M(x_1) \wedge M(x_2)& \text{ otherwise}
	\end{array}
	\right.\;.
\end{eqnarray*}
Now consider the randomized response mechanisms:
\begin{eqnarray*}
	M_{\rm RR}(x_i) &=& \left\{
	\begin{array}{rl}
		x_i & \text{with probability} \frac{\lambda}{1+\lambda}\;,\\
		\bar{x}_i & \text{with probability} \frac{1}{1+\lambda}\;.
	\end{array}
	\right.
\end{eqnarray*}

The dominance of this randomized response follows from a more general result proved in \cite{OV13}
which introduces a new operational interpretation of differential privacy mechanisms that
provides strong analytical tools to compare privacy mechanisms.

\begin{figure}[h]
	\begin{center}
	\includegraphics[width=.3\textwidth]{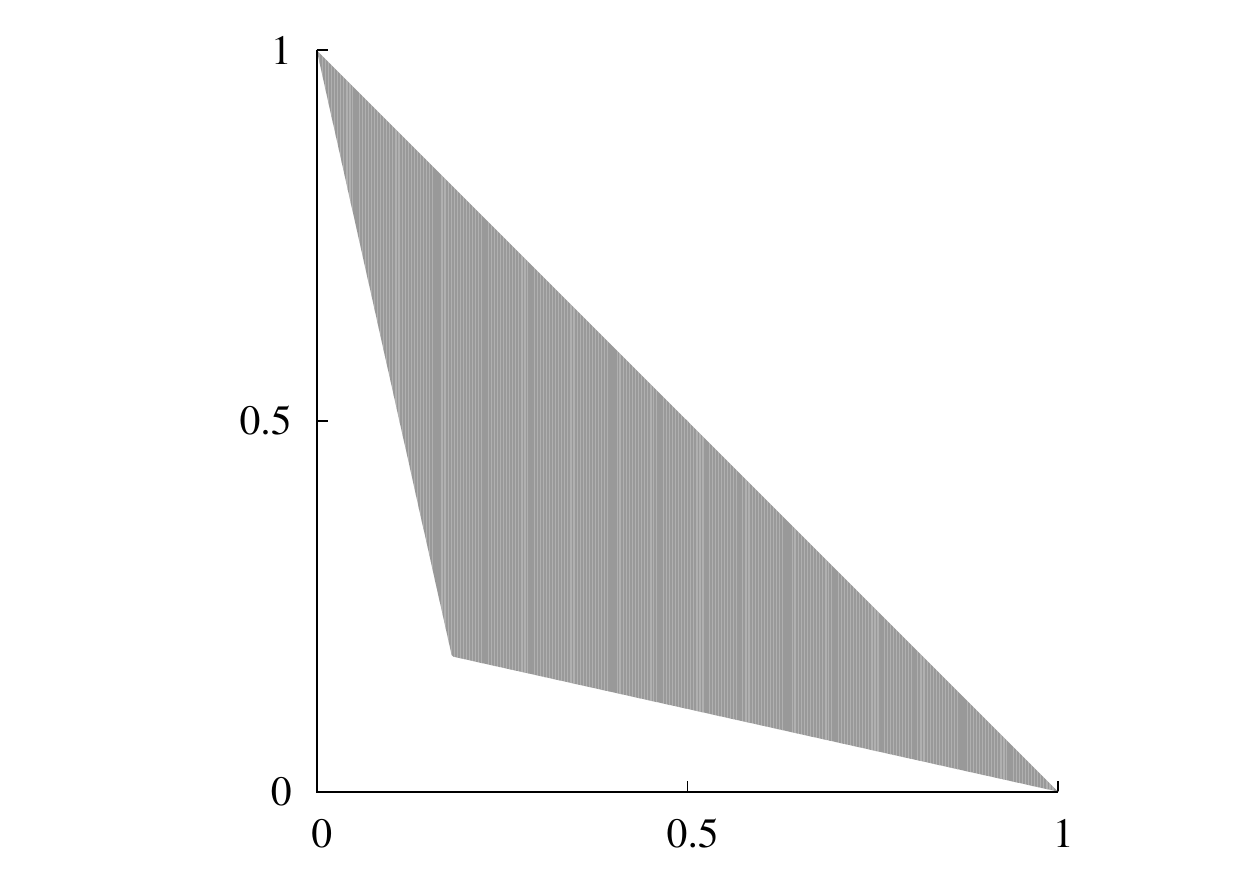}
	\put(-15,10){$P_{\rm MD}$}
	\put(-120,100){$P_{\rm FA}$}
	\end{center}
	\caption{Three regions $\cR_M$, $\cR_{M_{\rm RR}}$, and $\cR_\lambda$ are identical ($\varepsilon=1.5$).}
	\label{fig:region1}
\end{figure}

This crucially relies on the following representation of  the privacy
guarantees of a mechanism. Given a mechanism, consider a binary hypothesis
test on whether the original bit was a zero or a one based on the output of
the mechanism. Then, the two types of errors (false alarm and missed
detection) on this binary hypothesis testing problem defines a
two-dimensional region where one axis is $P_{\rm FA}$ and the other is
$P_{\rm MD}$. For a rejection set $S$ for rejecting the null hypothesis,
$P_{\rm FA}=\prob(M(x)\in S)$ and $P_{\rm MD}=\prob(M(x)\notin S)$. The
convex hull of the set of all pairs $(P_{\rm MD},P_{\rm FA})$ for all
rejection sets, define the hypothesis testing region. For example, the
mechanism $M$ corresponds to region $\cR_M$ and the randomized response
corresponds to region $\cR_{M_{\rm RR}}$ in Figure \ref{fig:region1}, which
happens to be identical.

Differential privacy conditions can be interpreted as  imposing a condition on
this region:
\begin{eqnarray*}
 	P_{\rm FA} + \lambda P_{\rm MD} &\geq & 1 \;,\;\;\text{ and } \;\;\;\;  \lambda P_{\rm FA} + P_{\rm MD}\;\geq\; 1\;,
\end{eqnarray*}
which defines a triangular region denoted by $\cR_\lambda$ and shown in
Figure~\ref{fig:region1}.
\begin{theorem}[{\cite[Theorem 2.3]{OV13}}]
	A mechanism is $\lambda$-differentially private if and only if
	the corresponding hypothesis testing region is included inside $\cR_\lambda$.
\end{theorem}
This is a special case of the original theorem which proves a more general theorem for $(\varepsilon,\delta)$-differential privacy.
We can immediately check that both $M$ and $M_{\rm RR}$ are $\lambda$-differentially private.

It is no coincidence that the regions $\cR_{M}$, $\cR_{M_{\rm RR}}$, and  $\cR_\lambda$
are identical.  It follows from the
next theorem on the  operational interpretation of differential privacy.
We say a mechanism $M_1$ {\em dominates} a mechanism $M_2$ if
$M_2(x)$ is conditionally independent of $x$ conditioned on $M_1(x)$. In
other words, we can construct the following Markov chain: $x-M_1(x)-M_2(x)$.
This is again equivalent to saying that there is another mechanism $T$ such
that $M_2(x) = T(M_1(x))$. Such an operational interpretation of differential
privacy brings both the natural  data processing inequality and the strong
converse to the data processing inequality, which follows from a celebrated
result of Blackwell on comparing two stochastic experiments \cite{Bla53}.
These inequalities, while simple by themselves, lead to surprisingly strong
technical results, and there is a long line of such a tradition in the
information theory literature: Chapter~17 of \cite{CT12} enumerates a
detailed list.

\begin{theorem}[{Data processing inequality for differential privacy \cite[Theorem 2.4]{OV13}}]
	If a mechanism $M_1$ dominates another mechanism $M_2$, then
	\begin{eqnarray*}
		\cR_{M_2} &\subseteq & \cR_{M_1} \;.
	\end{eqnarray*}
\end{theorem}

\begin{theorem}[{A strong converse to the data processing inequality \cite[Theorem 2.5]{OV13}}]
	For two mechanisms $M_1$ and $M_2$, 	there exists a coupling of the two mechanisms such that $M_1$ dominates $M_2$, if
	\begin{eqnarray*}
		\cR_{M_2} &\subseteq & \cR_{M_1} \;.
	\end{eqnarray*}
\end{theorem}
Among other things, this implies that among all $\lambda$-differentially private mechanisms,
the randomized response dominates all of them.
It follows that, for an arbitrary mechanism $M$,
there is another mechanism $T$ such that $M(x)=T(M_{\rm RR}(x))$.

In the two-party setting, this implies the desired claim that there is no
point in doing anything other than the randomized response, and that for the
AND example,  even though the protocol in \cite{GMPS} uses an alphabet of
three letters for each party, it is still able to achieve maximum accuracy,
because there is no reduction in the  hypothesis testing region. The final
decision is made as per $\hf(M(x_1),M(x_2))$. Without doing any calculations,
one could have guessed that this is achievable with randomized response which
uses only the minimal two letters
 by simply simulating $M(x_i)$ upon receiving $M_{\rm RR}(x_i)$,
namely, by computing $\hf_{\rm RR}(M_{\rm RR}(x_1),M_{\rm RR}(x_2)) =  \hf(T(M(x_1)),T(M(x_2)))$.
The new decision rule for randomized response is:
\begin{eqnarray*}
	\hf_{\rm RR}(M_{\rm RR}(x_1),M_{\rm RR}(x_2)) &=& \left\{
	\begin{array}{rl}
	0& \text{ if } (M_{\rm RR}(x_1),M_{\rm RR}(x_2))=(0,0)\\
	1 & \text{ if } (M_{\rm RR}(x_1),M_{\rm RR}(x_2))=(1,1)\\
	0 & \text{ if } (M_{\rm RR}(x_1),M_{\rm RR}(x_2))=(0,1)\text{ or } (1,0)\text{, then with probability }\frac{\lambda}{1+\lambda}\\
	1& \text{ if }(M_{\rm RR}(x_1),M_{\rm RR}(x_2))=(0,1)\text{ or } (1,0)\text{, then with probability }\frac{1}{1+\lambda}
	\end{array}
	\right. \;.
\end{eqnarray*}

\subsection{Multi-party  XOR computation}
\label{sec:XOR}

For a given function and a given accuracy measure,
analyzing the performance of the optimal protocol provides
the exact nature of the privacy-accuracy tradeoff.
Consider a scenario where a central observer wants to compute the XOR of all
the $k$-bits, each of which is $\lambda$-differentially private. In this
special case, we can apply our main theorem to analyze the accuracy exactly
in a combinatorial form, and we provide a proof in Section \ref{sec:xor}.
\begin{corollary}
	Consider $k$-party computation for $f_0(x)=x_1 \oplus \cdots \oplus x_k$, and the accuracy measure
	is one if correct and zero if not, i.e. $\w_0(0,0)=\w_0(1,1)=1$ and $\w_0(0,1)=\w_0(1,0)=0$.
	For any $\{\lambda\}$-differentially private protocol $P$ and any decision rule $\hf$,
	the average and worst-case accuracies are bounded by
	\begin{eqnarray*}
		\aveacc(P,w_0,f_0,\hf_0) &\leq&  \frac{\sum_{i=0}^{\lfloor k/2\rfloor} {k \choose 2i}\lambda^{k-2i} }{(1+\lambda)^k}\;,\;\;\text{ and }\;\;\;\;
		\wcacc(P,w_0,f_0\hf_0) \;\leq\;  \frac{\sum_{i=0}^{\lfloor k/2\rfloor} {k \choose 2i}\lambda^{k-2i} }{(1+\lambda)^k}\;,
	\end{eqnarray*}
	and the equality is achieved by the randomized response and optimal decision rules in \eqref{eq:scenariotwo1} and \eqref{eq:scenariotwo2}.	
\label{coro:xor}
\end{corollary}
The optimal decision for both  accuracies is simply
to output the XOR of the received privatized bits.
This is a strict generalization of a similar result in \cite{GMPS}, where
 XOR computation was studied but only for a two-party setting.
In the high privacy regime, where $\varepsilon\simeq0$ (equivalently $\lambda =e^\varepsilon \simeq 1$),
this implies that
\begin{eqnarray*}
	\aveacc = 0.5 + 2^{-(k+1)} \varepsilon^k + O(\varepsilon^{k+1})\;.
\end{eqnarray*}
The leading term is due to the fact that we are considering an accuracy measure of a Boolean function.
The second term of $2^{-(k+1)} \varepsilon^k$ captures the effect that,
we are essentially observing the XOR through $k$ consecutive binary symmetric channels with
flipping probability $\lambda/(1+\lambda)$.
Hence, the accuracy gets exponentially worse in $k$.
On the other hand, if those $k$-parties are allowed to collaborate, then they can compute the XOR in advance and only transmit the
privatized version of the XOR, achieving accuracy of $\lambda/(1+\lambda)=0.5+(1/4)\varepsilon^2 + O(\varepsilon^3)$.
This is always better than not collaborating, which is the bound in Corollary \ref{coro:xor}.

\section{Discussion}
\label{sec:discussion}
In this section, we discuss a few topics, 
each of which  are interesting but non-trivial to solve in any obvious way. 
Our main result is general and sharp, but we want to ask how we can push it even further. 

\begin{figure}[h]
	\begin{center}
	\includegraphics[width=.45\textwidth]{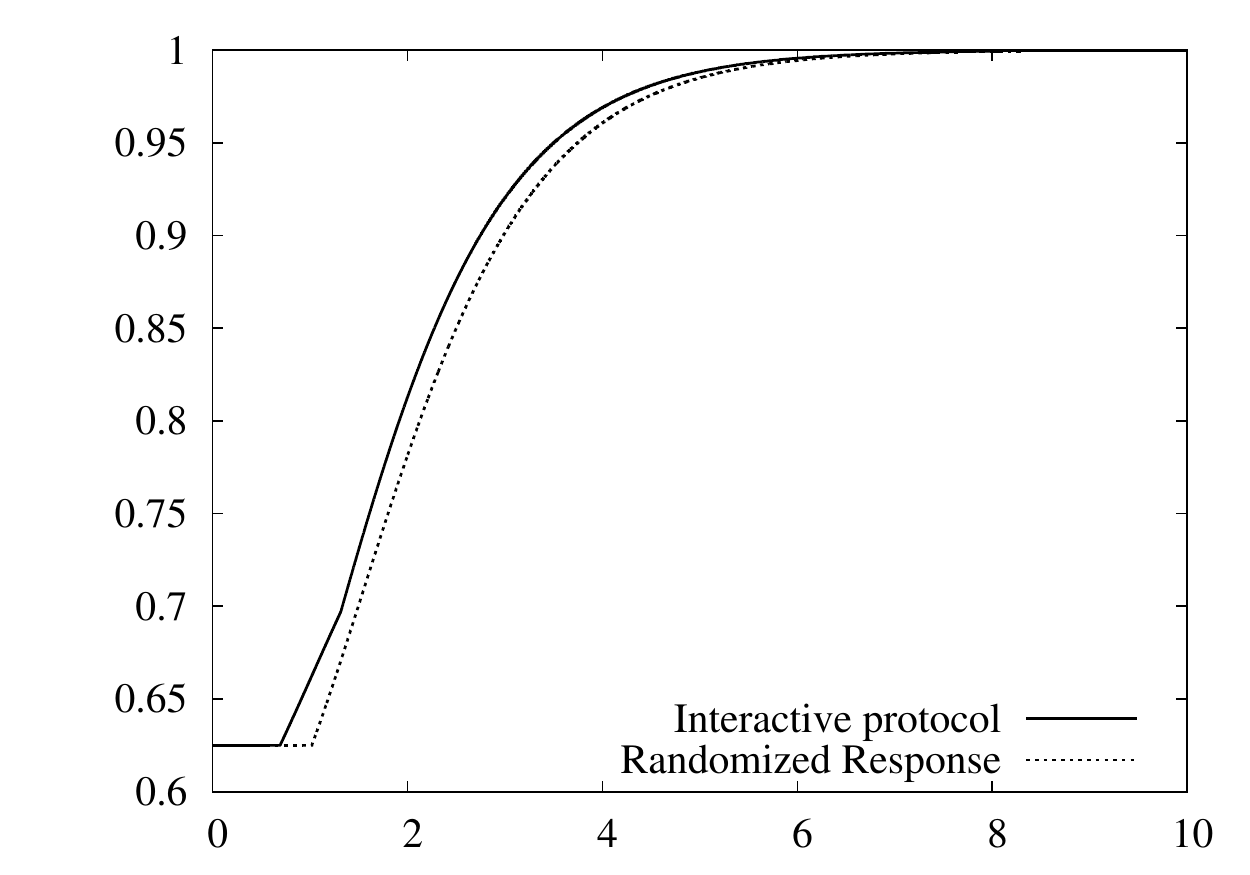}
	\put(-127,-5){Privacy level $\varepsilon$}
	\put(-205,145){Average accuracy}
	\includegraphics[width=.45\textwidth]{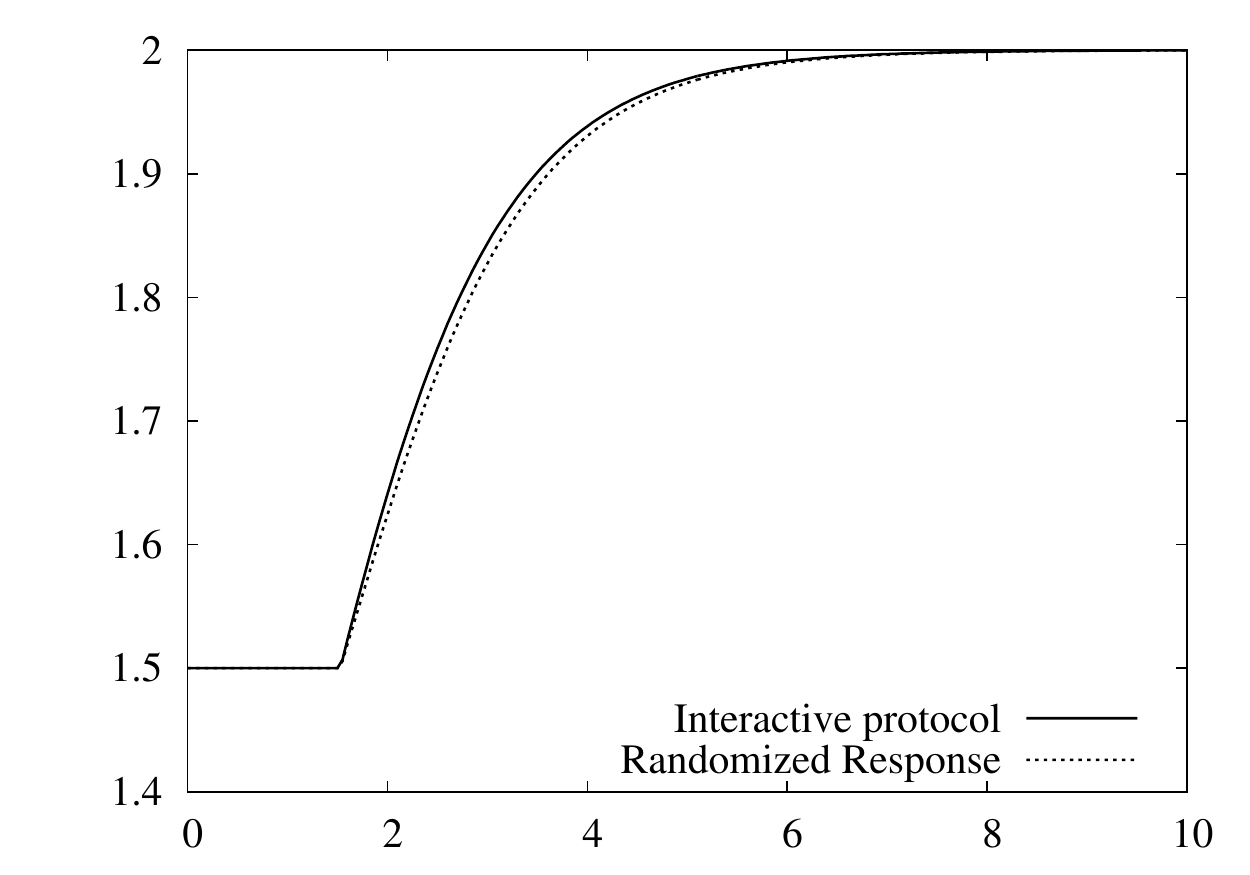}
	\put(-127,-5){Privacy level $\varepsilon$}
	\put(-205,145){Average accuracy}
	\end{center}
	\caption{Interactive protocols can improve over the randomized response, when each party owns multiple bits,  
	for computing XOR or AND (left) and computing the Hamming distance (right).}
	\label{fig:compare}
\end{figure}

\bigskip
\noindent{\bf Generalization to multiple bits.}
When each party owns multiple bits, 
it is possible that interactive protocols improve over the randomized response protocol. 
For example, consider the first party with one bit $x$ and the second party has two bits $y_1$ and $y_2$. 
Each bit needs to be protected as per $\varepsilon$-differential privacy. 
A central observer wishes to compute the following function: 
\begin{eqnarray*}
	f(x,y_1,y_2) &=& \left\{ 
		\begin{array}{rl}
			y_1 \oplus \; y_2 & \text{ if } x=0 \;,\\
			y_1 \wedge \; y_2 & \text{ if } x=1\;.
		\end{array}
	\right.
\end{eqnarray*}
Randomized response would publish privatized versions of $x$, $y_1$, and $y_2$ according to \eqref{eq:rr}.  
In an interactive scheme, 
looking at $\tilde{x}$, the second party publishes (the privatized version of) either $y_1 \oplus y_2$ (if $\tilde{x} = 0$) 
or $y_1\wedge y_2$ (if $\tilde{x} = 1$). 
Upon receiving the privatized data, the central observer makes optimal decisions in each case. 
Figure \ref{fig:compare} illustrates how these two protocols compare in terms of average accuracy, 
where the accuracy is one if the approximation is correct and zero if the approximation is incorrect.  
For $\varepsilon=0$, both protocols cannot do better than 
the best random guess of zero. which achieves average accuracy of $5/8=0.625$.
For large $\varepsilon$, both protocols achieve the best accuracy of one.

Another example of multiple bit multi-party computation is  studied in \cite{MMPT}. 
There are two parties each owning two bits of data $x\in\{0,1\}^2$ and $y\in\{0,1\}^2$, 
and a third party wants to compute the Hamming distance $d_H(x,y) = \sum_{i=1}^2 |x_i-y_i|$. 
Assuming each bit needs to be protected, 
the randomized response would reveal each bit via Equation \ref{eq:rr}. 
On the other hand, we can design an interactive scheme where 
one party reveals its two bits via the randomized response, and 
the other party then outputs its best estimate of the Hamming distance obeying differential privacy guarantees, i.e. 
\begin{eqnarray*}
	\prob\big(\, \tilde{y} |\tx=(0,0),y=(0,0)\,\big) \;=\;\left\{ 
	\begin{array}{rl}
	 	 \frac{\lambda^2}{\lambda(1+\lambda)} & \text{ for } \tilde{y}=0\\
	 	 \frac{\lambda - 1}{\lambda(1+\lambda)} & \text{ for } \tilde{y}=1\\
	 	 \frac{1}{\lambda(1+\lambda)} & \text{ for } \tilde{y}=2
	\end{array}
	\right.\\
	\prob\big(\, \tilde{y} |\tx=(0,0),y=(0,1)\text{ or }(1,0)\,\big) \;=\;\left\{ 
	\begin{array}{rl}
	 	 \frac{\lambda}{\lambda(1+\lambda)} & \text{ for } \tilde{y}=0\\
	 	 \frac{\lambda ^2-\lambda}{\lambda(1+\lambda)} & \text{ for } \tilde{y}=1\\
	 	 \frac{\lambda}{\lambda(1+\lambda)} & \text{ for } \tilde{y}=2
	\end{array}
	\right.\\
	\prob\big(\, \tilde{y} |\tx=(0,0),y=(1,1)\,\big) \;=\;\left\{ 
	\begin{array}{rl}
	 	 \frac{1}{\lambda(1+\lambda)} & \text{ for } \tilde{y}=0\\
	 	 \frac{\lambda -1}{\lambda(1+\lambda)} & \text{ for } \tilde{y}=1\\
	 	 \frac{\lambda^2}{\lambda(1+\lambda)} & \text{ for } \tilde{y}=2
	\end{array}
	\right.
\end{eqnarray*}
where $\tx\in\{0,1\}^2$ is the output of the first party via randomized response, 
and $\tilde{y}\in\{0,1,2\}$ is the output of the second party. 
Figure \ref{fig:compare} illustrates how these two protocols compare in terms of average accuracy, 
where the accuracy is $2-|d_H(x,y)-\hat{d}|$ where $\hat{d}$ is the optimal decision made by the third party. 

one if the approximation is correct and zero if the approximation is incorrect.  
For $\varepsilon=0$, both protocols cannot do better than 
the best random guess of zero. which achieves average accuracy of $5/8=0.625$.
For large $\varepsilon$, both protocols achieve the best accuracy of one.

\bigskip\noindent{\bf Approximate differential privacy.}
A common generalization of differential privacy,
 known as the approximate differential privacy, 
 is to allow a small slack of $\delta\geq0$ in the privacy condition\cite{Dwo06,DMNS06}. 
In the multi-party context, a protocol $P$ is $(\varepsilon_i,\delta_i)$-differentially private for the $i$-th party if for 
all $i\in[k]$, and all $x_i,x_i'\in\{0,1\}$, $x_{-i}\in\{0,1\}^{k-1}$, and for all subset $T \subseteq \cT$, 
\begin{eqnarray}
	\prob(\tau \in T |x_i,x_{-i}) &\leq& e^{\varepsilon_i} \prob(\tau\in T |x_i',x_{-i})  + \delta_i \;.
	\label{eq:appxdp}
\end{eqnarray} 
It is natural to ask if the linear programming (LP) approach presented in this paper can be 
extended to identify the optimal multi-party protocol under $\{(\varepsilon_i,\delta_i)\}$-differential privacy. 
The LP formulations of \eqref{eq:primal-lp} and \eqref{eq:primal-lp-wc} heavily rely on the fact that 
any differentially private protocol $P$ can be decomposed as the combination of the matrix $S$ and the $\theta^{(y)}$'s. 
Since the differential privacy constraints are invariant under scaling of $P^{(y)}_{\tau}$, 
one can represent the scale-free pattern of the distribution with $S_\tau$
and the scaling with $\theta_\tau^{(y)}$. 
This is no longer true for $\{(\varepsilon_i,\delta_i)\}$-differential privacy, and the analysis technique does not generalize. 

\bigskip\noindent{\bf Correlated sources.}
When the data $x_i$'s are correlated (e.g. each party observe a noisy version of the state of the world), 
knowing $x_i$ reveals some information on other parties' bits. 
In general, revealing correlated data requires careful coordination between multiple parties. 
The analysis techniques developed in this paper 
does not generalize to correlated data, since the crucial rank-one tensor structure of $S_\tau^{(y)}$ is no longer present.



\bigskip\noindent{\bf Extensions to general utility functions.}
A surprising aspect of the main result is that even though the worst-case accuracy is a concave function over the protocol $P$, 
the maximum is achieved at an extremal point of the manifold of rank-1 tensors. 
This suggests that there is a deeper geometric structure of the problem, leading to 
possible universal optimality of the randomized response for a broader class of utility functions. 
It is an interesting task to understand the geometric structure of the problem, and to ask 
what class of utility functions lead to optimality of the randomized response.

\section{Acknowledgement}
The authors gratefully acknowledge the useful discussions with O.~Pandey and
M.~Prabhakaran.

\newpage
\bibliographystyle{plain}
\bibliography{privacy,references}

\begin{thebibliography}{10}

\bibitem{BNO08}
Amos Beimel, Kobbi Nissim, and Eran Omri.
\newblock Distributed private data analysis: Simultaneously solving how and
  what.
\newblock In {\em Advances in Cryptology--CRYPTO 2008}, pages 451--468.
  Springer, 2008.

\bibitem{ben1988completeness}
Michael Ben-Or, Shafi Goldwasser, and Avi Wigderson.
\newblock Completeness theorems for non-cryptographic fault-tolerant
  distributed computation.
\newblock In {\em Proceedings of the twentieth annual ACM symposium on Theory
  of computing}, pages 1--10. ACM, 1988.

\bibitem{Bla53}
D.~Blackwell.
\newblock Equivalent comparisons of experiments.
\newblock {\em The annals of mathematical statistics}, 24(2):265--272, 1953.

\bibitem{BDMN05}
A.~Blum, C.~Dwork, F.~McSherry, and K.~Nissim.
\newblock Practical privacy: the sulq framework.
\newblock In {\em Proceedings of the twenty-fourth ACM SIGMOD-SIGACT-SIGART
  symposium on Principles of database systems}, pages 128--138. ACM, 2005.

\bibitem{BN10}
Hai Brenner and Kobbi Nissim.
\newblock Impossibility of differentially private universally optimal
  mechanisms.
\newblock In {\em Foundations of Computer Science (FOCS), 2010 51st Annual IEEE
  Symposium on}, pages 71--80. IEEE, 2010.

\bibitem{CKN11}
J.~A. Calandrino, A.~Kilzer, A.~Narayanan, E.~W. Felten, and V.~Shmatikov.
\newblock " you might also like:" privacy risks of collaborative filtering.
\newblock In {\em Security and Privacy (SP), 2011 IEEE Symposium on}, pages
  231--246. IEEE, 2011.

\bibitem{CSS12}
K.~Chaudhuri, A.~Sarwate, and K.~Sinha.
\newblock Near-optimal differentially private principal components.
\newblock In {\em Advances in Neural Information Processing Systems}, pages
  989--997, 2012.

\bibitem{CSS13}
K.~Chaudhuri, A.~D. Sarwate, and K.~Sinha.
\newblock A near-optimal algorithm for differentially-private principal
  components.
\newblock {\em Journal of Machine Learning Research}, 14:2905--2943, 2013.

\bibitem{CMS11}
Kamalika Chaudhuri, Claire Monteleoni, and Anand~D Sarwate.
\newblock Differentially private empirical risk minimization.
\newblock {\em The Journal of Machine Learning Research}, 12:1069--1109, 2011.

\bibitem{Chaum88}
David Chaum, Claude Cr{\'e}peau, and Ivan Damgard.
\newblock Multiparty unconditionally secure protocols.
\newblock In {\em Proceedings of the twentieth annual ACM symposium on Theory
  of computing}, pages 11--19. ACM, 1988.

\bibitem{CT12}
T.~M. Cover and J.~A. Thomas.
\newblock {\em Elements of information theory}.
\newblock John Wiley \& Sons, 2012.

\bibitem{DJW13}
J.~C. Duchi, M.~I. Jordan, and M.~J. Wainwright.
\newblock Local privacy and statistical minimax rates.
\newblock In {\em Foundations of Computer Science (FOCS), 2013 IEEE 54th Annual
  Symposium on}, pages 429--438. IEEE, 2013.

\bibitem{Dwo06}
C.~Dwork.
\newblock Differential privacy.
\newblock In {\em Automata, languages and programming}, pages 1--12. Springer,
  2006.

\bibitem{DMNS06}
C.~Dwork, F.~McSherry, K.~Nissim, and A.~Smith.
\newblock Calibrating noise to sensitivity in private data analysis.
\newblock In {\em Theory of Cryptography}, pages 265--284. Springer, 2006.

\bibitem{Dwo11}
Cynthia Dwork.
\newblock Differential privacy: A survey of results.
\newblock In {\em Theory and Applications of Models of Computation}, pages
  1--19. Springer, 2008.

\bibitem{DKMMN}
Cynthia Dwork, Krishnaram Kenthapadi, Frank McSherry, Ilya Mironov, and Moni
  Naor.
\newblock Our data, ourselves: Privacy via distributed noise generation.
\newblock In {\em Advances in Cryptology-EUROCRYPT 2006}, pages 486--503.
  Springer, 2006.

\bibitem{GV12}
Quan Geng and Pramod Viswanath.
\newblock The optimal mechanism in differential privacy.
\newblock {\em arXiv preprint arXiv:1212.1186}, 2012.

\bibitem{GV13}
Quan Geng and Pramod Viswanath.
\newblock The optimal mechanism in differential privacy: Multidimensional
  setting.
\newblock {\em arXiv preprint arXiv:1312.0655}, 2013.

\bibitem{GRS09}
A.~Ghosh, T.~Roughgarden, and M.~Sundararajan.
\newblock Universally utility-maximizing privacy mechanisms.
\newblock {\em SIAM Journal on Computing}, 41(6):1673--1693, 2012.

\bibitem{GMW87}
O.~Goldreich, S.~Micali, and A.~Wigderson.
\newblock How to play any mental game.
\newblock In {\em Proceedings of the Nineteenth Annual ACM Symposium on Theory
  of Computing}, STOC '87, pages 218--229, New York, NY, USA, 1987. ACM.

\bibitem{GMPS}
Vipul Goyal, Ilya Mironov, Omkant Pandey, and Amit Sahai.
\newblock Accuracy-privacy tradeoffs for two-party differentially private
  protocols.
\newblock In {\em Advances in Cryptology--CRYPTO 2013}, pages 298--315.
  Springer, 2013.

\bibitem{GS10}
Mangesh Gupte and Mukund Sundararajan.
\newblock Universally optimal privacy mechanisms for minimax agents.
\newblock In {\em Proceedings of the twenty-ninth ACM SIGMOD-SIGACT-SIGART
  symposium on Principles of database systems}, pages 135--146. ACM, 2010.

\bibitem{HR12}
M.~Hardt and A.~Roth.
\newblock Beating randomized response on incoherent matrices.
\newblock In {\em Proceedings of the forty-fourth annual ACM symposium on
  Theory of computing}, pages 1255--1268. ACM, 2012.

\bibitem{HSR08}
N.~Homer, S.~Szelinger, M.~Redman, D.~Duggan, W.~Tembe, J.~Muehling, J.~V.
  Pearson, D.~A. Stephan, S.~F. Nelson, and D.~W. Craig.
\newblock Resolving individuals contributing trace amounts of dna to highly
  complex mixtures using high-density snp genotyping microarrays.
\newblock {\em PLoS genetics}, 4(8):e1000167, 2008.

\bibitem{KOV14}
P.~Kairouz, S.~Oh, and P.~Viswanath.
\newblock Extremal mechanisms for local differential privacy.
\newblock In {\em Advances in neural information processing systems}, 2014.

\bibitem{KT13}
M.~Kapralov and K.~Talwar.
\newblock On differentially private low rank approximation.
\newblock In {\em Proceedings of the Twenty-Fourth Annual ACM-SIAM Symposium on
  Discrete Algorithms}, pages 1395--1414. SIAM, 2013.

\bibitem{KLNRS11}
Shiva~Prasad Kasiviswanathan, Homin~K Lee, Kobbi Nissim, Sofya Raskhodnikova,
  and Adam Smith.
\newblock What can we learn privately?
\newblock {\em SIAM Journal on Computing}, 40(3):793--826, 2011.

\bibitem{Kil00}
Joe Kilian.
\newblock More general completeness theorems for secure two-party computation.
\newblock In {\em Proceedings of the thirty-second annual ACM symposium on
  Theory of computing}, pages 316--324. ACM, 2000.

\bibitem{KQR09}
Robin K{\"u}nzler, J{\"o}rn M{\"u}ller-Quade, and Dominik Raub.
\newblock Secure computability of functions in the it setting with dishonest
  majority and applications to long-term security.
\newblock In {\em Theory of Cryptography}, pages 238--255. Springer, 2009.

\bibitem{KN06}
E.~Kushilevitz and N.~Nisan.
\newblock {\em Communication Complexity}.
\newblock Cambridge University Press, 2006.

\bibitem{MMPT}
Andrew McGregor, Ilya Mironov, Toniann Pitassi, Omer Reingold, Kunal Talwar,
  and Salil Vadhan.
\newblock The limits of two-party differential privacy.
\newblock In {\em Foundations of Computer Science (FOCS), 2010 51st Annual IEEE
  Symposium on}, pages 81--90. IEEE, 2010.

\bibitem{MT07}
F.~McSherry and K.~Talwar.
\newblock Mechanism design via differential privacy.
\newblock In {\em Foundations of Computer Science, 2007. FOCS'07. 48th Annual
  IEEE Symposium on}, pages 94--103. IEEE, 2007.

\bibitem{NS08}
A.~Narayanan and V.~Shmatikov.
\newblock Robust de-anonymization of large sparse datasets.
\newblock In {\em Security and Privacy, 2008. SP 2008. IEEE Symposium on},
  pages 111--125. IEEE, 2008.

\bibitem{OV13}
Sewoong Oh and Pramod Viswanath.
\newblock The composition theorem for differential privacy.
\newblock {\em arXiv preprint arXiv:1311.0776}, 2013.

\bibitem{PP12}
Manoj~M Prabhakaran and Vinod~M Prabhakaran.
\newblock On secure multiparty sampling for more than two parties.
\newblock In {\em Information Theory Workshop (ITW), 2012 IEEE}, pages 99--103.
  IEEE, 2012.

\bibitem{RFP10}
Benjamin Recht, Maryam Fazel, and Pablo~A Parrilo.
\newblock Guaranteed minimum-rank solutions of linear matrix equations via
  nuclear norm minimization.
\newblock {\em SIAM review}, 52(3):471--501, 2010.

\bibitem{Swe97}
L.~Sweeney.
\newblock Weaving technology and policy together to maintain confidentiality.
\newblock {\em The Journal of Law, Medicine \& Ethics}, 25(2-3):98--110, 1997.

\bibitem{War65}
S.~L. Warner.
\newblock Randomized response: A survey technique for eliminating evasive
  answer bias.
\newblock {\em Journal of the American Statistical Association},
  60(309):63--69, 1965.

\bibitem{Yao82}
Andrew~C Yao.
\newblock Protocols for secure computations.
\newblock In {\em 2013 IEEE 54th Annual Symposium on Foundations of Computer
  Science}, pages 160--164. IEEE, 1982.

\end{thebibliography}

\newpage
\appendix

\section{Appendix}

\subsection{Proof of Corollary \ref{coro:xor}}
\label{sec:xor}
Let $\tX$ denote the random output of the randomized response,
and let $f(\tX)$ denote the XOR of all $k$ bits.
Notice that
$P(X,\tX)=(\lambda^{k-d_h(X,\tX)})/(1+\lambda)^k$ where $d_h(\cdot,\cdot)$ denotes the Hamming distance.
For a given $\tX$ the decision is either $f(\tX)$ or the complement of it.
We will first show that $f(\tX)$ is the optimal decision rule.

It is sufficient to show that $\E[w(f(X),f(\tX))|\tX] \geq \E[w(f(X),\bar{f}(\tX))|\tX]$.
Since, $\E[w(f(X),f(\tX))|\tX] = \sum_{i\text{ even}} {k \choose i} \lambda^{k-i} /(1+\lambda)^k$
and $\E[w(f(X),\bar{f}(\tX))|\tX] = \sum_{i\text{ odd}} {k \choose i} \lambda^{k-i} /(1+\lambda)^k$,
it follows that
$$\E[w(f(X),f(\tX))|\tX] - \E[w(f(X),\bar{f}(\tX))|\tX] = (\lambda-1)^k / (1+\lambda)^k \geq 0\;,$$
since $\lambda \geq 1$. By symmetry, the decision rule is the same for all $\tX$, and also for the worst case accuracy.
This finishes the desired characterization of the optimal accuracy.

To get the asymptotic analysis of the accuracy,
notice that $\E[w(f(X),f(\tX))] + \E[w(f(X),\bar{f}(\tX))] = 1$
and $\E[w(f(X),f(\tX))] + \E[w(f(X),\bar{f}(\tX))] = (\lambda-1)^k/(1+\lambda)^k = (e^\varepsilon-1)^k/(2+(e^\varepsilon-1))^k= (1/2)^k\varepsilon^k + O(\varepsilon^{k+1})$.
It follows that
$\E[w(f(X),f(\tX))] = 1/2 + (1/2)^{k+1}\varepsilon^k + O(\varepsilon^{k+1})$.

\subsection{Proof of Lemma \ref{lem:polytope}}
\label{sec:polytope}
Consider the following half space for $\reals^{2^k}$.
For an $a\in\{-1,+1\}^k$, the half space $H_a$ is defined as the set of $T\in\reals^{2^k}$ satisfying
\begin{eqnarray}
   (-1)^k\,\Big( \prod_{j\in[k]} a_j \Big) \; \sum_{x\in\{0,1\}^k} \Big( T_x \prod_{i\in[k]} (-\lambda_i)^{a_i\,{x}_i } \Big)\;  \geq \;0 \;.
\label{eq:Ha}
\end{eqnarray}
We claim that
\begin{eqnarray*}
	\cP_{\{\lambda_i\}} &=&\left\{  T\in \bigcap_{a\in\{-1,+1\}^k} H_a \;\Big|\; T_{0\ldots0}=1 \right\} \;.
\end{eqnarray*}
It is straightforward to see that $\cM_{\{\lambda_i\}}$ is inside the intersection of all $2^k$ half-spaces:
all tensors in $\cM_{\{\lambda_i\}}$ satisfy
\begin{eqnarray*}
	   (-1)^k\,\Big( \prod_{j\in[k]} a_j \Big) \;\prod_{i\in[k]} \Big(1-\lambda^{a_i}t_i\Big) \geq 0\;,
\end{eqnarray*}
for all $a\in\{-1,+1\}^k$. This immediately implies that the tensors satisfy \eqref{eq:Ha}.
To show that it is indeed the convex hull, we need to show that $\cM_{\{\lambda_i\}}$ intersects with the boundary of $\cP_{\{\lambda_i\}}$
at every corner point.
$\cP_{\{\lambda_i\}}$ as defined above is $2^k-1$ dimensional  polytope in $2^k$ dimensional space,
with at most $2^k$ faces and $2^k$ corner points.
Each corner point is an intersection of $2^k-1$ half spaces and the one hyperplane defined by $T_{0\ldots0}=1$.

Consider a corner point of $\cM_{\{\lambda_i\}}$ represented by $a\in\{-1,+1\}^k$ as
\begin{eqnarray*}
	T^{(a)} = [1\,,\,\lambda_1^{a_1} ]\otimes \cdots \otimes[1\,,\,\lambda_k^{a_k}]\;.
\end{eqnarray*}
It follows that $T^{(a)}$ is an intersection of $2^k-1$ half spaces $H_b$ for $b\neq a$.
Hence, every corner point of $\cP_{\{\lambda_i\}}$ intersects with $\cM_{\{\lambda_i\}}$.
This finishes the proof.

\end{document}